\documentclass[graybox]{svmult}

  \usepackage[cmex10]{amsmath}

\usepackage{mathptmx}       
\usepackage{helvet}         
\usepackage{courier}        
\usepackage{type1cm}        
\usepackage{makeidx}         
\usepackage{graphicx}        
\usepackage{multicol}        
\usepackage[bottom]{footmisc}
\usepackage{mathrsfs}
\usepackage{cleveref}

\makeindex             
                       
                       \usepackage{amssymb}
                       \DeclareMathOperator*{\cov}{Cov} 
                       \DeclareMathOperator*{\var}{Var}	
                       \DeclareMathOperator*{\bias}{Bias}	
                       \DeclareMathOperator*{\mse}{MSE}	
                       \usepackage{dsfont} 
                       \usepackage{paralist} 
                       
                     \newtheorem{assumption}{Assumption}

   						\usepackage[ruled,vlined]{algorithm2e}

                       \usepackage{enumerate} 
                       
                       \usepackage[outdir=./]{epstopdf}
                       \usepackage{caption}
                       \usepackage{subcaption}
                       \newcommand{\estIP}{I^b}  
                       \newcommand{\estRW}{T^b_{RW}}  
                       \newcommand{\estFN}{T^b_{FN}}   
                       \newcommand{\estUN}{T^b_{UN}} 
                       \newcommand{\truevalue}{\bar{\rho}}   
                       \newcommand{\samplingbudget}{b}   

                       \newcommand{\st}{x}
                       
                       \newcommand{\obs}{y}

                       \newcommand{\varmi}{H^{-}_{n}}
                       \newcommand{\varp}{H_{n}}
                       \newcommand{\gn}{K_{n}}
                       \newcommand{\rn}{\mathbf{R}_{n}}
                       
                       \newcommand{\vertexnum}{M}
                       
                       \newcommand{\Vertexset}{V}

                       \newcommand{\degdist}{P}
                       \newcommand{\nodem}{m}

                       \newcommand{\bst}{\st}

                       \newcommand{\Ndeg}{D}

                       \newcommand{\dtime}{n}
                       \newcommand{\sample}{\nu}
                       \newcommand{\degg}{k}
                       \newcommand{\seq}{l}
                       
                       \newcommand{\beq}{\begin{equation}}
                       	\newcommand{\eeq}{\end{equation}}
                       \newcommand{\nodeobs}{\hat{\bst}}

                       \newcommand{\steady}{\pi}
                       \newcommand{\weight}{W}

                       \newlength{\dhatheight}

\begin{document}

\title*{Information Diffusion in Social Networks: Friendship Paradox based Models and Statistical Inference}
\author{Vikram Krishnamurthy and Buddhika Nettasinghe}
\institute{Vikram Krishnamurthy \at Cornell Tech and School of Electrical \& Computer Engineering, Cornell University\newline \email{vikramk@cornell.edu}
\and Buddhika Nettasinghe \at Cornell Tech and School of Electrical \& Computer Engineering, Cornell University\newline \email{dwn26@cornell.edu}
\and This research was supported by the Army Research Office under grant W911NF-17-1-0335 and National Science Foundation under grant 1714180.}
%
%
\maketitle


\abstract{Dynamic models and statistical inference for the diffusion of information in social networks is an area which has witnessed remarkable progress in the last decade due to the proliferation of social networks. Modeling and inference of diffusion of information has applications in targeted advertising and marketing, forecasting elections, predicting investor sentiment and identifying epidemic outbreaks. This chapter discusses three important aspects related to information diffusion in social networks: (i)~How does observation bias named \textit{friendship paradox} (a graph theoretic consequence) and \textit{monophilic contagion} (influence of friends of friends) affect information diffusion dynamics. (ii)~How can social networks adapt their structural connectivity depending on the state of information diffusion. (iii)~How one can estimate the state of the network induced by information diffusion. The motivation for all three topics considered in this chapter stems from recent findings in network science and social sensing. Further, several directions for future research that arise from these topics are also discussed.}

\section{Introduction}
\label{sec:introduction}
Information diffusion refers to how the opinions (states) of individual nodes in a social network (graph) evolve with time. The two phenomena that give rise to information diffusion in social networks are 1~-~contagion and 2~-~homophily. Contagion-based diffusions are driven by influence of neighbors whereas homophily-based diffusions are driven by properties of nodes (which are correlated among neighbors)~\cite{aral2009distinguishing,mcpherson2001birds,shalizi2011homophily}. Dynamic models and statistical inference for such information diffusion processes in social networks (such as news, innovations, cultural fads, etc) has witnessed remarkable progress in the last decade due to the proliferation of social media networks such as Facebook, Twitter, Youtube, Snapchat and also online reputation systems such as Yelp and Tripadvisor. Models and inference methods for information diffusion in social networks are useful in a wide range of applications including selecting influential individuals for targeted advertising and marketing \cite{kempe2003, nettasinghe2018influence,seeman2013}, localization of natural disasters \cite{sakaki2010earthquake}, forecasting elections \cite{nettasinghe2018your} and predicting sentiment of investors in financial markets \cite{pang2008opinion,bollen2011twitter}. For example, \cite{asur2010predicting} shows that models based on the rate of Tweets for a particular product can outperform market-based prediction methods.

This chapter deals with the contagion-based information diffusion in large scale social networks. In such contagion-based information diffusion (henceforth referred to as information diffusion) processes, states (which could represent opinions, voting intentions, purchase of a product, etc.) of individuals in the network evolve over time as a probabilistic function of the states of their neighbors. Popular models for studying information diffusion processes over networks include Susceptible-Infected (SI), Susceptible-Infected-Susceptible~(SIS), Susceptible-Infected-Recovered (SIR) and Susceptible-Exposed-Infected-Recovered (SEIR) \cite{hethcote2000mathematics,easley2010networks}. Apart from these models, several recent works also investigated information diffusions using real-world social network datasets: \cite{romero2011differences} studied the spread of hashtags on Twitter, \cite{bakshy2012role} conducted larges scale field experiments to identify the causal effects of peer influence in information diffusion, \cite{lerman2010information} studied how the network structure affects dynamics of information flow using Digg and Twitter datasets to track how interest in new stories spread over them. 


\begin{figure*}[t!]
	\centering
	\includegraphics[width=4.5in ]{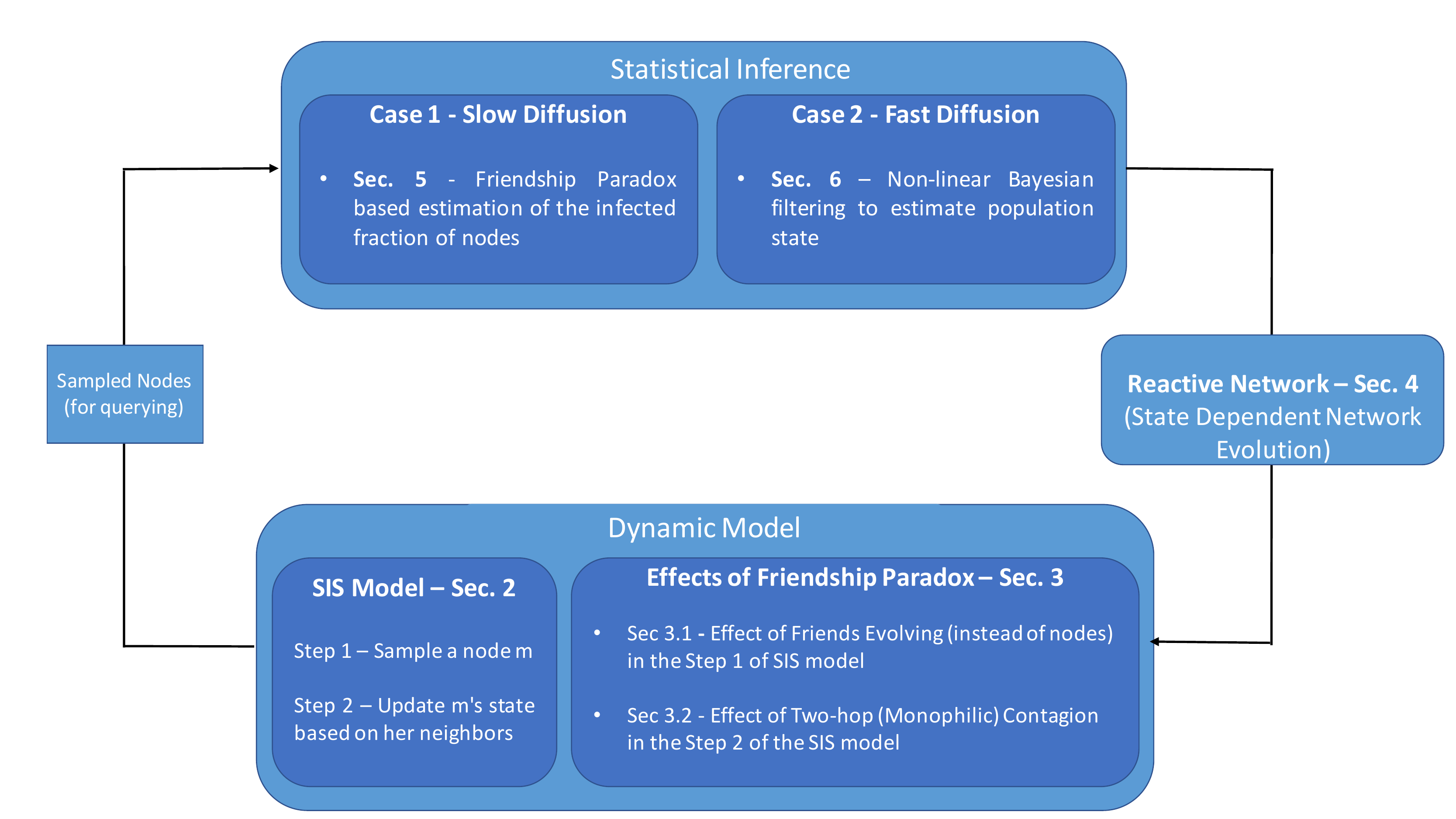}
	\caption{Block diagram illustrating the main topics covered and their organization in this chapter. The main topics are unified under the central theme of dynamic modeling and statistical inference of information diffusion processes over social networks.}
	\label{fig:block_diagram}
\end{figure*}

\subsection*{Main Topics and Organization}
\label{subsec:main_results_organization}

In this chapter, we consider a discrete time version of the SIS model on an undirected network which involves two steps (detailed in Sec. \ref{sec:preliminaries}) at each time instant. In the first step, a randomly sampled individual (agent) $m$ from the population observes $d(m)$ (degree of $m$) number of randomly selected agents (neighbors of $m$). In the second step, based on the $d(m)$ observations, the state of agent $m$ evolves probabilistically to one of the two possible states: infected or susceptible. 

 In the context of this discrete time SIS model, next, we briefly discuss the main topics studied in this chapter, motivation for studying them and how they are organized throughout this chapter. Further, how the main topics discussed in different sections are interconnected with each other and unified under the main theme of dynamic modeling and statistical inference of information diffusion processes is illustrated in Fig. \ref{fig:block_diagram}.
\bigskip

\noindent
1. {\bf Friendship Paradox based Variants of the SIS model}
\newline
The first topic (Sec. \ref{sec:effects_FP_SISModel}) considered in this chapter is the effects of friendship paradox on the SIS model. \textit{Friendship paradox} refers to a graph theoretic consequence that was introduced in 1991 by Scott. L. Feld in \cite{feld1991}. 
Feld's original statement of the friendship paradox is ``on average, the number of friends of a random friend is always greater than or equal to the number of friends of a random individual". Here, a random friend refers to a random end node $Y$ of a randomly chosen edge (a pair of friends). This statement is formally stated in Theorem \ref{th:friendship_paradox} below. 

\begin{theorem}[Friendship Paradox \cite{feld1991}]
	\label{th:friendship_paradox}
	Let ${G = (V,E)}$ be an undirected graph, $X$ be a node chosen uniformly from $V$ and, $Y$ be a uniformly chosen node from a uniformly chosen edge $e\in E$. Then,
	\begin{equation}
	\mathbb{E} \{d(Y)\} \geq \mathbb{E}\{d(X)\},
	\end{equation} where, $d(X)$ denotes the degree of $X$. 
	
\end{theorem}

Studying the friendship paradox (Theorem \ref{th:friendship_paradox}) based variants of the SIS model is motivated by the following two assumptions made in most works (for example, see \cite{lopez2008diffusion,jackson2007relating,lopez2012influence,pastor2001epidemic,jackson2007diffusion}) related to SIS models. 
\begin{compactenum}[i.]
	\item \label{assumption:uniform_m} Each node is equally likely to update her state at each time instant i.e. uniform nodes are sampled in the first step of the SIS model. 
	
	\item \label{assumption:immediate_nbr} Individuals decide whether to get infected or not based only on their (immediate) neighbors' states.
\end{compactenum}

In real world social networks (e.g. Facebook or Twitter), the frequency with which a node updates her state (e.g. opinion) depends on the number of her social interactions (i.e. degree) according to recent findings \cite{hodas2012visibility}. This contradicts the assumption~\ref{assumption:uniform_m}. As a solution, Sec. \ref{subsec:effects_step1} studies the modified SIS model where the state of a random friend (instead of a random node) evolves at each time instant. This modification to the standard SIS model reflects the fact that high degree nodes evolve more often in real world social networks. The main result of Sec. \ref{subsec:effects_step1} shows that this modification results in different dynamics (compared to the standard SIS model) but, with the same critical thresholds (which determine if the information diffusion process will eventually die away or not) on the parameters of the SIS model.

Further, it has been shown in several recent works (e.g. \cite{altenburger2018monophily,fatouhi2018conjoining}) that the individuals' attributes and decisions in real world social networks are affected by two-hop neighbors (i.e. friends of friends). This two-hop neighbors' effects in real world networks are ignored in the assumption \ref{assumption:immediate_nbr} of the standard SIS model. As an alternative, Sec. \ref{subsec:critical_thersholds} considers the case where friends of friends influences the state evolutions instead of friends. We refer to this two-hop influence as \textit{monophilic contagion} since the correlation between two-hop nodes is called \textit{monophily}\footnote{The concept of monophily presented in \cite{altenburger2018monophily} does not give a causal interpretation but only the correlation between two-hop neighbors of an undirected graph. What we consider is monophilic contagion (motivated by monophily): the information diffusion caused by the influence of two hop neighbors in an undirected network.} \cite{altenburger2018monophily}. Main result of Sec.~\ref{subsec:critical_thersholds} shows that information diffusion processes under monophilic contagion (decision to adopt a product, an idea, etc. is based on two-hop neighbors) spreads more easily (i.e. has a smaller critical threshold) compared to information diffusion under non-monophilic contagion (one-hop influence) as a result of the friendship paradox\footnote{ Effects of the friendship paradox on information diffusion have been considered in \cite{lattanzi2015,bagrow2017friends, lee2018impact}. However, the effect of friendship paradox on information diffusion under monophilic contagion (two-hop influence) has not been explored in the literature to the best of our knowledge.}. This result also suggests that talking to random friends of friends could be more efficient (compared to talking to random friends) in spreading rumors, news, etc. The well known friendship paradox based immunization approach \cite{cohen2003efficient} that immunizes random friends (instead of random nodes) relies on a similar argument: random friends have larger degrees (compared to random nodes) and are more critical to the spreading of a disease.
\bigskip

\noindent
2. {\bf SIS Model and Reactive Networks: Collective Dynamics}
\newline
Modeling a network as a deterministic graph does not capture information diffusion processes in real world networks. Several works proposed and analyzed evolving graph models: \cite{ogura2016epidemic} studied the adaptive susceptible-infected-susceptible (ASIS) model where susceptible individuals are allowed to temporarily cut edges connecting them to infected nodes in order to prevent the spread of the infection, \cite{pare2018epidemic} analyzed the stability of epidemic processes over time-varying networks and provides sufficient conditions for convergence, \cite{paarporn2017networked} studied a SIS process over a static contact network where the nodes have partial information about the epidemic state and react by limiting their interactions with their neighbors when they believe the epidemic is currently prevalent. These serve as the motivation for Sec. \ref{sec:active_networks} where the underlying network is modeled as a \textit{reactive network}: a random graph process whose transition probabilities at each time instant depend on the state of the  information diffusion process. The main result of Sec. \ref{sec:active_networks} shows that, when the network is a \textit{reactive network} which randomly evolves depending on the state of the information diffusion, the collective dynamics of the network and the diffusion process can be approximated (under some assumptions) by an ordinary differential equation (ODE) with an algebraic constraint. From a statistical modeling and machine learning perspective, the importance of this result relies on the fact that it provides a simple deterministic approximation of the collective stochastic dynamics of a complex system (an SIS process on a random graph, both evolving on the same time scale).
\bigskip

\noindent
3. {\bf Estimating the Population State under Slow and Fast Information Diffusion}
\newline
Sec. \ref{sec:FP_based_polling} and Sec. \ref{sec: NL_Filtering} deal with estimating the population states induced by the SIS model under two cases: 
\begin{enumerate}
	\item the information diffusion is slow and hence states of nodes can be treated as fixed for the purpose of the estimating the population state 
	
	\item the information diffusion is fast and hence, states of nodes cannot be treated as fixed for the purpose of estimation.
\end{enumerate}

\bigskip
\noindent
{\bf Case 1 - Polling under slow information diffusion}
\newline
Polling is the method of asking a question from randomly (according to some distribution) sampled individuals and averaging their responses \cite{graefe2014accuracy}. Therefore, the accuracy of a poll depends on two factors: (i) - method of sampling respondents for the poll (ii) - question presented to the sampled individuals. For example, when forecasting the outcome of an election, asking people “Who do you think will win?" (expectation polling) is better compared to “Who will you vote for?" (intent polling) \cite{rothschild2011}. This is due to the fact that an individual will name the candidate that is most popular among her friends in expectation polling (and thus summarizing a number of individuals in the social network) instead of providing her own voting intention. Motivated by such polling approaches, Sec. \ref{sec:FP_based_polling} presents two friendship paradox based polling algorithms that aim to estimate the fraction of infected individuals by querying random friends instead of random individuals. Since random friends have more friends (and hence, have more observations) than random individuals on average, the proposed methods yield a better (in a mean squared error sense) estimate compared to intent polling as well as expectation polling with random nodes.

\bigskip
\noindent
{\bf Case 2 - Bayesian filtering under fast information diffusion}
\newline
Friendship paradox based polling algorithms in Sec. \ref{sec:FP_based_polling} assume that information diffusion takes place on a slower time scale compared to the time taken to poll the individuals. Hence, for the purpose of the polling algorithm, the states of the individuals can be treated as fixed. However, such approaches are not applicable in situations where the information diffusion takes place on the same time scale as the time scale on which individuals are polled i.e. cases where measurement (polling) process takes place on the same time scale as the one on which the information spreads. Further, the information diffusion process constitute a non-linear (in the states) dynamical system as we show subsequently in Sec. \ref{sec:preliminaries}. Hence applying optimal filtering algorithms such as Kalman filter is not possible. These facts motivate the non-linear filtering algorithm discussed in Sec. \ref{sec: NL_Filtering} which recursively (with each new measurement) computes the conditional mean of the state of the information diffusion (given the observations). 
\bigskip

\noindent
{\bf Summary}
\newline
The main topics explored in this chapter bring together two important aspects related to a stochastic dynamical system mentioned at the beginning of this chapter: dynamic modeling and statistical inference. In terms of the dynamic modeling aspect (which is covered in Sec. Sec. \ref{sec:preliminaries}, Sec. \ref{sec:effects_FP_SISModel} and Sec. \ref{sec:active_networks}), we are interested in understanding how changes to the standard SIS-model can result in different dynamics and stationary states. In terms of statistical inference (covered in Sec. \ref{sec:FP_based_polling} and Sec. \ref{sec: NL_Filtering}), we are interested in estimating the underlying state of the population induced by the model. Fig. \ref{fig:block_diagram} illustrates how these topics are organized in this chapter and are interconnected under the unifying theme.  Rather than delving into detailed proofs, our aim in this chapter is to stress several novel insights.

\section{Mean-Field Dynamics of SIS Model and Friendship Paradox}
\label{sec:preliminaries}

Mean-Field dynamics refers to a simplified model of a (stochastic) system where the stochastic dynamics are replaced by deterministic dynamics.
Much of this research is based on the seminal work of Kurtz \cite{kurtz1981approximation} on population dynamics models.
In this section, we first discuss how mean-field dynamics can be used as a deterministic model of a SIS diffusion process over an undirected network. Since an SIS diffusion over a social network is a Markov process whose state space grows exponentially with the number of individuals, mean-field dynamics offers a deterministic model that is analytically tractable \cite{lopez2008diffusion,lopez2012influence,jackson2007relating,krishnamurthy2017tracking}. Then, several recent generalizations of the original version of the friendship paradox are presented. The purpose of mean-field dynamics and the friendship paradox results discussed in this section is to study (in Sec. \ref{sec:effects_FP_SISModel}) how friendship paradox based changes to the standard SIS model (e.g. random friends evolving instead of random nodes in the step 1 of SIS model) can result in different mean-field dynamics and critical thresholds.


\subsection{Discrete time SIS Model}
\label{subsec:SIS_model}

Consider a social network represented by an undirected graph $G = (V, E)$ where $V = \{1, 2, \dots, M\}$ denotes the set of nodes.  At each discrete time instant $n$, a node $v\in V$ of the network can take the state $s_n^{(v)} \in \{0,1 \}$ where, $0$ denotes the susceptible state and $1$ denotes the infected state. The degree $d(v) \in \{1,\dots, D\}$ of a node $v \in V$  is the number of nodes connected to $v$ and, $M(k)$ denotes the total number of nodes with degree $k$. Then, the degree distribution $P(k) = \frac{M(k)}{M}$ is the probability that a randomly selected node has degree $k$. Further, we also define the population state $\bar{x}_n(k)$ as the fraction of nodes with degree $k$ that are infected (state $1$) at time $n$ i.e.
\begin{equation}
	\label{eq:population_state}
	\bar{x}_n(k) = \frac{1}{M(k)}{\sum_{v \in V}\mathds{1}_{\{d(v) = k,\,s_n^{(v)} = 1 \}}}, \quad k = 1,\dots ,D.
\end{equation}

For this setting, we adopt the SIS model used in \cite{krishnamurthy2017tracking, krishnamurthy2014interactive} which is as follows briefly. 

\vspace{0.1cm}
\noindent
{\bf Discrete Time SIS Model: }At each discrete time instant $n$,
\begin{compactenum}
	\item[\bf Step 1:] A node $m \in V$ is chosen with uniform probability $p^{X} (m) = 1/M$ where, $M$ is the number of nodes in the graph. 
	
	\item[\bf Step 2:] The state $s_n^{(v)} \in \{0,1 \}$ of the sampled node $m$ (in Step 1) evolves to $s_{n+1}^{(v)} \in \{0,1 \}$ with transition probabilities that depend on the degree of $m$, number of infected neighbors of $m$, population state of the network $\bar{x}_n$\footnote{$\bar{x}_n(k)$ is the fraction of infected nodes with degree $k$ i.e. $\bar{x}_n(k) = \frac{M^1(k)}{M(k)}$ where $M^1(k)$ is the number of infected nodes with degree $k$ and $M(k)$ is the number of nodes with degree $k$.} and the current state of $s^{(m)}_n$. 
\end{compactenum}

\vspace{0.25cm}
Note that the above model is a Markov chain with a state space consisting of $2^M$ states (since each of the $M$ nodes can be either infected or susceptible at any time instant). Due to this exponentially large state space, the discrete time SIS model is not mathematically tractable. However, we are interested only in the fraction of the infected nodes (as opposed to the exact state out of the $2^M$ states) and therefore, it is sufficient to focus on the dynamics of the population state $\bar{x}_n$ defined in (\ref{eq:population_state}) instead of the exact state of the infection.

\subsection{Mean-Field Dynamics Model}

Mean-field dynamics has been used in literature (e.g. \cite{lopez2008diffusion,lopez2012influence,jackson2007relating,kurtz1981approximation,krishnamurthy2017tracking}) as a useful means of obtaining a tractable deterministic model of the dynamics of the population state $\bar{x}_n$.  The following result from \cite{krishnamurthy2017tracking} shows how mean-field dynamics model closely approximates the stochastic dynamics of the  true population state $\bar{x}_{n}$.

\begin{theorem}[Mean-Field Dynamics]
	\label{th:MFD}
	\begin{compactenum}
		\item The population state defined in (\ref{eq:population_state}) evolves according to the following stochastic difference equation driven by martingale difference process:      
		\begin{align}
			\label{eq:martingale_representation}
			\bar{x}_{n+1}(k) &= \bar{x}_{n}(k) + \frac{1}{M}[P_{01}(k, \bar{x}_n) - P_{10}(k, \bar{x}_n)] +\zeta_n \\
			\text{where,}\nonumber\\
			P_{01}(k, \bar{x}_n) &= (1-\bar{x}_n(k)) \mathbb{P}(s_{n+1}^{m} = 1 \vert s_{n}^{m} = 0, d(m) = k, \bar{x}_n ) \label{eq:scaled_transition_prob_1}\\
			P_{10}(k, \bar{x}_n)  &= \bar{x}_n(k)\mathbb{P}(s_{n+1}^{m} = 0 \vert s_{n}^{m} = 1, d(m) = k, \bar{x}_n ).\label{eq:scaled_transition_prob_2}
		\end{align} are the scaled transition probabilities of the states and, $\zeta_n$ is a martingale difference process with $\vert \vert \zeta_n\vert \vert_2 \leq \frac{\Gamma}{M}$ for some positive constant $\Gamma$.

		\item Consider the mean-field dynamics process associated with the population state:
		\begin{equation}
			\label{eq:MFD_approximation_X}
			x_{n+1} (k) = x_{n}(k) + \frac{1}{M}\big(	P_{01}(k, x_n) - P_{10}(k, x_n)	\big)
		\end{equation}
		where, $P_{01}(k, x_n)$ and $P_{10}(k, x_n)$ are as defined in (\ref{eq:scaled_transition_prob_1}), (\ref{eq:scaled_transition_prob_2}) and $x_0 = \bar{x}_0$.
		Then, for a time horizon of T points, the deviation between the mean-field dynamics (\ref{eq:MFD_approximation_X})  and the actual population state $\bar{x}_n$ of the SIS model satisfies
		\begin{equation}
			\mathbb{P}\{\underset{0\leq n \leq T}{\max}\vert\vert  x_n -\bar{x}_n   \vert\vert_{\infty} \geq \epsilon   \} \leq C_1\exp(-C_2 \epsilon^2M)
		\end{equation} for some positive constants $C_1, C_2$ providing $T=O(M)$.
	\end{compactenum}
\end{theorem}

First part of Theorem \ref{th:MFD} is the martingale representation of a Markov chain (which is the population state $\bar{x}_n$). Note from (\ref{eq:martingale_representation}) that the dynamics of the population state $\bar{x}_n$ resemble a stochastic approximation recursion (new state is the old state plus a noisy term). Hence, the trajectory of the population state $\bar{x}_n$ should converge (weakly) to the deterministic trajectory given by the ODE corresponding to the mean-field dynamics in (\ref{eq:MFD_approximation_X}) as the size of the network $M$ goes to infinity i.e. the step size of the stochastic approximation algorithm goes to zero (for details, see \cite{krishnamurthy2016, kushner2003}). Second part of the Theorem \ref{th:MFD} provides an exponential bound on the deviation of the mean-field dynamics model from the actual population state for a finite length of the sample path. In the subsequent sections of this chapter, the mean-field approximation (\ref{eq:MFD_approximation_X}) is utilized to explore the topics outlined in Sec. \ref{sec:introduction}.

\subsection{Friendship Paradox}
\label{subsec:friendship_paradox}
Recall that the original version of friendship paradox (Theorem \ref{th:friendship_paradox}) is a comparison between the average degrees of a random individual $X$ and a random friend $Y$. This subsection reviews recent generalizations and extensions of the original version of friendship paradox stated in Theorem \ref{th:friendship_paradox}.

The original version of the friendship paradox (Theorem \ref{th:friendship_paradox}) can be described more generally in terms of likelihood ratio ordering as follows:
\begin{theorem}[Friendship Paradox - Version 1 \cite{cao2016}]
	\label{th:friendship_paradox_v1_cao}
	Let ${G = (V,E)}$ be an undirected graph, $X$ be a node chosen uniformly from $V$ and, $Y$ be a uniformly chosen node from a uniformly chosen edge $e\in E$. Then,
	\begin{equation}
	d(Y) \geq_{lr} d(X),
	\end{equation} where, $\geq_{lr}$ denotes the likelihood ratio dominance\footnote{\label{fn:lr}A discrete random variable $Y$ (with a probability mass function $f_Y$) likelihood ratio dominates a discrete random variable $X$ (with a probability mass function $f_X$), denoted $Y \geq_{lr} X$ if, ${f_Y(n)/ f_X(n)}$ is an increasing function of $n$. Further, likelihood ratio dominance implies larger mean. Therefore, Theorem \ref{th:friendship_paradox_v1_cao} implies that $\mathbb{E} \{d(Y)\} \geq \mathbb{E}\{d(X)\}$ as stated in Theorem \ref{th:friendship_paradox}.}. 
	
\end{theorem} 

Theorem \ref{th:friendship_paradox_v2_cao} (based on \cite{cao2016}) states that a similar result holds when the degrees of a random node $X$ and a random friend $Z$ of a random node $X$ are compared as well.

\begin{theorem}[Friendship Paradox - Version 2 \cite{cao2016}]
	\label{th:friendship_paradox_v2_cao}
	Let ${G = (V,E)}$ be an undirected graph, $X$ be a node chosen uniformly from $V$ and, $Z$ be a uniformly chosen neighbor of a uniformly  chosen node from $V$. Then, 
	\begin{equation}
		\label{eq:fosd}
		d(Z)\geq_{fosd} d(X)
	\end{equation} 
	where, $\geq_{fosd}$ denotes the first order stochastic dominance\footnote{\label{fn:fosd}A discrete random variable $Y$ (with a cumulative distribution function $F_Y$) first order stochastically dominates a discrete random variable $X$ (with a cumulative distribution function $F_X$), denoted $Y \geq_{fosd} X$ if, ${F_Y(n) \leq F_X(n)}$, for all $n$. Further, first order stochastic dominance implies larger mean. Hence, Theorem \ref{th:friendship_paradox_v2_cao} implies that $\mathbb{E} \{d(Z)\} \geq \mathbb{E}\{d(X)\}$.}. 
\end{theorem}

The intuition behind the two versions of the friendship paradox (Theorems \ref{th:friendship_paradox_v1_cao} and \ref{th:friendship_paradox_v2_cao}) stems from the fact that individuals with a large number of friends (high degree nodes) appear as the friends of a large number of individuals. Therefore, high degree nodes contributes to an increase in the average number of friends of friends. On the other hand, individuals with smaller number of friends appear as friends of a smaller number of individuals. Hence, they do not cause a significant change in the average number of friends of friends. 

Friendship paradox, which in essence is a sampling bias observed in undirected social networks has gained attention as a useful tool for estimation and detection problems in social networks. For example, \cite{eom2015tail} proposes to utilize friendship paradox as a sampling method for reduced variance estimation of a heavy-tailed degree distribution, \cite{christakis2010,garcia2014using, sun2014efficient} explore how the friendship paradox can be used for detecting a contagious outbreak quickly, \cite{seeman2013, lattanzi2015,horel2015scalable, kim2015social, kumar2018network} utilizes friendship paradox for maximizing influence in a social network, \cite{nettasinghe2018your} proposes friendship paradox based algorithms for efficiently polling a social network (e.g. to forecast an election) in a social network, \cite{jackson2016friendship} studies how the friendship paradox in a game theoretic setting can systematically bias the individual perceptions. 

Several generalizations, extensions and consequences of friendship paradox have also been proposed in the literature. \cite{eom2014} shows how friendship paradox can be generalized to other attributes (apart from the degree) such as income and happiness when there exists a positive correlation between the attribute and the degree. Related to this work, \cite{higham2018centrality} showed that certain other graph based centrality measures such as eigenvector centrality and Katz centrality (under certain assumptions) exhibit a version of the friendship paradox, leading to the statement ``your friends are more important than you, on average". \cite{hodas2013} extended the concept of friendship paradox to directed networks and empirically showed that four versions of friendship paradox which compare the expected in- and out- degrees of random friends and random followers to expected degree of a random node can exist in directed social networks such as Twitter. \cite{lerman2016} discusses ``majority illusion" an observation bias that stems from friendship paradox which makes many individuals in a social network to observe that a majority of their neighbors are in a particular state (e.g. possesses an iPhone), even when that state is globally rare. Similarly, \cite{bagrow2017friends, kramer2016multistep, bollen2017happiness, fotouhi2014generalized} also discuss various other generalizations and consequences of friendship paradox.

\section{Effects of Friendship Paradox on SIS Model}
\label{sec:effects_FP_SISModel}

Sec. \ref{sec:preliminaries} reviewed the discrete time SIS model that involves two steps and, showed how mean-field dynamics can be used as a deterministic model of an SIS information diffusion process. In the context of the SIS model, the aim of this section is to explore how changes (motivated by examples discussed in Sec. \ref{subsec:main_results_organization}) to the first step (sampling a node $m$) and the second step ($m$ updates its state probabilistically based on the states of neighbors) of the standard SIS model are reflected in the deterministic mean-field dynamics model and its critical threshold. The changes to the standard SIS model (Sec. \ref{subsec:SIS_model}) that we explore are motivated by friendship paradox in the sense that we consider 1 - random friends (instead of random nodes) are sampled in the first step, 2 - state of the sampled node is updated based on the states of friends of friends (instead of immediate friends).

\subsection{Effect of the Sampling Distribution in the Step 1 of the SIS Model }
\label{subsec:effects_step1}
Recall from Sec. \ref{subsec:friendship_paradox} that we distinguished between three sampling methods for a network ${G = (V, E)}$: a random node $X$, a random friend $Y$ and, a random friend $Z$ of a random node. Further, recall that in the discrete-time SIS model explained in Sec. \ref{subsec:SIS_model}, the node $m$ that whose state evolves is sampled uniformly from $V$ i.e. $m\overset{d}{=} X$. This section studies the effect of random friends ($Y$ or $Z$) evolving at each time instant instead of random nodes ($X$) i.e. the cases where ${m\overset{d}{=} Y}$ or ${m\overset{d}{=} Z}$. Following is the main result in this section:
\begin{theorem}
	\label{thm:MFD_samping_YZ}
	Consider the discrete time SIS model presented in Sec. \ref{subsec:SIS_model}.
	\begin{compactenum}
		\item If the node $m$ is a random end  $Y$ of random link i.e. node $m$ with degree $d(m)$ is chosen with probability ${p^Y(m) = \frac{d(m)}{\sum_{v\in V}d(v)}}$, then the stochastic dynamics of the SIS model can be approximated by,
		\begin{equation}
		\label{eq:MFD_sampling_Y}
		x_{n+1} (k) = x_{n}(k) + \frac{1}{M} \frac{k}{\bar{k}}\big(	P_{01}(k, x_n) - P_{10}(k, x_n)	\big),
		\end{equation}
		where $\bar{k}$ is the average degree of the graph $G = (V, E)$. 
		
		\item If the node $m$ is a random neighbor $Z$ of a random node $X$, then the stochastic dynamics of the SIS model can be approximated by,
		\begin{align}
		x_{n+1} (k) &= x_{n}(k) +\frac{1}{M} \bigg({\sum_{k'}\frac{P(k)}{P(k')}P(k\vert k')}\bigg) \big(	P_{01}(k, x_n) - P_{10}(k, x_n)	\big), 	\label{eq:MFD_sampling_Z}
		\end{align}
		where $\bar{k}$ is the average degree of the graph $G = (V, E)$, $P$ is the degree distribution and $P(k\vert k')$ is the probability that a random neighbor of a degree $k'$ node is of degree $k$.  Further, if the network is a degree-uncorrelated network i.e. $P(k\vert k')$ does not depend on $k'$, then (\ref{eq:MFD_sampling_Z}) will be the same as $(\ref{eq:MFD_sampling_Y})$. 
	\end{compactenum}
\end{theorem}

Theorem \ref{thm:MFD_samping_YZ}  is proved in \cite{nettasinghe2018contagions}. Theorem \ref{thm:MFD_samping_YZ} shows that, if the node $m$ sampled in the step 1 of the SIS model (explained in Sec. \ref{subsec:SIS_model}), is chosen to be a random friend or a random friend of a random node, then different elements $x_n(k)$ of the mean-field approximation evolves at different rates. This result allows us to model the dynamics of the population state in the more involved case where, frequency of the evolution of an individual is proportional his/her degree (part 1 - e.g. high degree nodes change opinions more frequently due to higher exposure) and also depends on the degree correlation (part 2 - e.g. nodes being connected to other similar/different degree nodes changes the frequency of changing the opinion).

\vspace{0.25cm}
\begin{remark} [Invariance of the critical thresholds to the sampling distribution in step 1] 
	\normalfont
	The stationary condition for the mean-field dynamics is obtained by setting $x_{n+1}(k) - x_{n}(k) = 0$ for all $k \geq 1$. Comparing (\ref{eq:MFD_approximation_X}) with (\ref{eq:MFD_sampling_Y}) and (\ref{eq:MFD_sampling_Z}), it can be seen that this condition yields the same expression $P_{01}(k, x_n) - P_{10}(k, x_n) = 0$, for all three sampling methods (random node - $X$, random end of a random link $Y$ and, a random neighbor $Z$ of a random node). Hence, the critical thresholds of the SIS model are invariant to the distribution from which the node $m$ is sampled in step 1. This leads us to Sec. \ref{subsec:critical_thersholds} where, modifications to the step 2 of the SIS model are analyzed in terms of the critical thresholds. 
\end{remark}

\subsection{Critical Thresholds for Unbiased-degree Networks}
\label{subsec:critical_thersholds}
In Sec. \ref{subsec:effects_step1} of this paper, we focused on step 1 of the SIS model (namely, distribution with which the node $m$ is drawn at each time instant) and, showed that different sampling methods for selecting the node $m$ result in different mean-field dynamics with the same stationary conditions. In contrast, the focus of this subsection is on the step 2 of the SIS model (namely, the probabilistic evolution of the state of the node $m$ sampled in step 1) and, how changes to step 2 would result in different stationary conditions and critical thresholds. More specifically, we are interested in understanding the effects on the SIS information diffusion process caused by \textit{monophilic contagion}: node $m$'s state evolves based on the states of random friends of friends (two-hop neighbors). This should be contrasted to the standard SIS information diffusion processes based on non-monophilic contagion where, evolution of node $m$'s state is based on states of random friends (one-hop neighbors).

\subsubsection{Critical Thresholds of Information Diffusion Process under Monophilic and Non-Monophilic Contagion Rules}

Recall the SIS model reviewed in Sec. \ref{subsec:SIS_model} again. We limit our attention to the case of \textit{unbiased-degree} networks and viral adoption rules discussed in \cite{lopez2016overview}. 

\vspace{0.25cm}
\noindent
{\bf Unbiased-degree network:} In an unbiased-degree network, neighbors of agent $m$ sampled in the step 1 of the SIS model are $d(m)$ (degree of agent $m$) number of uniformly sampled agents (similar in distribution to the random variable $X$) from the network. Therefore, in an unbiased-degree network, any agent is equally likely to be a neighbor of the sampled (in the step 1 of the SIS model) agent $m$.

\vspace{0.25cm}
\noindent
{\bf Viral adoption rules\footnote{	The two rules (case 1 and case 2) are called viral adoption rules as they consider the total number of infected nodes (denoted by $a^X_m$ and $a^Z_m$ in case 1 and case 2 respectively) in the sample in contrast to the persuasive adoption rules that consider the fraction of infected nodes in the sample \cite{lopez2012influence}.}:}  If the sampled agent $m$ (in the step 1 of the SIS model) is an infected agent, she becomes susceptible with a constant probability $\delta$.  If the sampled agent $m$ (in the step 1 of the SIS model) is a susceptible (state $0$) agent, she samples $d(m)$ (degree of $m$) number of other agents $X_1, X_2, \dots, X_{d(m)}$ (neighbors of $m$ in the unbiased-degree network) from the network and, updates her state (infected or susceptible) based on one of the following rules:
\begin{compactenum}
	\item [ Case 1 - Non-monophilic contagion:] For each sampled neighbor $X_i$, $m$ observes the state of $X_i$. Hence, agent $m$ observes the states of $d(m)$ number of random nodes. Let $a^X_m$ denote the number of infected agents among $X_1,\dots, X_{d(m)}$. Then, the susceptible agent $m$ becomes infected with probability $\nu\frac{a^X_m}{D}$ where, $0 \leq \nu\leq 1$ is a constant and $D$ is the largest degree of the network.
	
	\vspace{0.2cm}
	\item[ Case 2 - Monophilic contagion:] For each sampled neighbor $X_i$, $m$ observes the state of a random friend $Z_i\in \mathcal{N}(X_i)$ of that neighbor. Hence, agent $m$ observes the states of $d(m)$ number of random friends $Z_1,\dots, Z_{d(m)}$ of random nodes $X_1,\dots,X_{d(m)}$. Let $a^Z_m$ be the number of infected agents among $Z_1, \dots, Z_{d(m)}$. Then, the susceptible agent $m$ becomes infected with probability $\nu\frac{a^Z_m}{D}$ where, $0 \leq \nu\leq 1$ is a constant and $D$ is the largest degree of the network.
\end{compactenum}

In order to compare the non-monophilic and monophilic contagion rules, we look at the conditions on the model parameters for which, each rule leads to a positive fraction of infected nodes starting from a small fraction of infected nodes i.e. a positive stationary solution to the mean-field dynamics (\ref{eq:MFD_approximation_X}). The main result is the following (proof given in \cite{nettasinghe2018contagions}):
\begin{theorem}
	\label{thm:observe_XZ}
	Consider the SIS model described in Sec. \ref{subsec:SIS_model}. Define the effective spreading rate as $\lambda = \frac{\nu}{\delta}$ and let $X$ be a random node and $Z$ be a random friend of $X$.
	\begin{compactenum}
		\item Under the non-monophilic contagion rule (Case 1), the mean-field dynamics equation ($\ref{eq:MFD_approximation_X}$) takes the form,
		\begin{align}
		\label{eq:MFD_approximation_X_observe_X}
		x_{n+1} (k) &= x_{n}(k) + \frac{1}{M}\big(	(1-{x}_n(k))\frac{\nu k \theta_n^X}{D}- x_n(k)\delta\big)\\
		\quad \text{where},\nonumber\\
		\theta_n^X &= \sum_{k}P(k) x_n(k) 
		\end{align} is the probability that a randomly chosen node $X$ at time $n$ is infected. Further, there exists a positive stationary solution to the mean field dynamics (\ref{eq:MFD_approximation_X_observe_X}) for case 1 if and only if
		\begin{equation}
		\label{eq:diff_threshold_X}
		\lambda > \frac{D}{\mathbb{E}\{d(X)\}} = \lambda^*_X
		\end{equation}
		
		\item Under the monophilic contagion rule (Case 2), the mean-field dynamics equation (\ref{eq:MFD_approximation_X}) takes the form,
		\begin{align}
		\label{eq:MFD_approximation_X_observe_Z}
		x_{n+1} (k) &= x_{n}(k) + \frac{1}{M}\big(	(1-{x}_n(k))\frac{\nu k \theta_n^Z}{D}- x_n(k)\delta	\big)\\
		\text{where,}\nonumber\\
		\theta_n^Z &= \sum_{k}\bigg(\sum_{k'}P(k')P(k|k')\bigg) x_n(k)
		\end{align} is the probability that a randomly chosen friend $Z$ of a randomly chosen node $X$ at time $n$ is infected\footnote{\label{fn:jdd}We use $P(k|k')$ to denote the conditional probability that a node with degree $k'$ is connected to a node with degree $k$. More specifically $P(k|k') = \frac{e(k,k')}{q(k)}$ where $e(k,k')$ is the joint degree distribution of the network and $q(k)$ is the marginal distribution that gives the probability of random end (denoted by random variable $Y$ in Theorem \ref{th:friendship_paradox}) of random link having degree $k$. We also use $\sigma_q$ to denote the variance of $q(k)$ in subsequent sections.}. Further, there exists a positive stationary solution to the mean field dynamics (\ref{eq:MFD_approximation_X_observe_Z}) if and only if
		\begin{equation}
		\label{eq:diff_threshold_Z}
		\lambda > \frac{D}{\mathbb{E}\{d(Z)\}}= \lambda^*_Z
		\end{equation}
	\end{compactenum}
\end{theorem}

The infection spreading under the monophilic contagion rule (Case 2 of Theorem \ref{thm:observe_XZ}) can also be thought of as representing the network by the square graph (corresponding to the square of the adjacency matrix of the original network). Proceeding that way would also yield the same critical threshold as in the Case 2 of Theorem \ref{thm:observe_XZ}. Theorem \ref{thm:observe_XZ} allows us to analyze the effects of friendship paradox and degree-assortativity on the contagion process as discussed in the next subsection.

\subsubsection{Effects of Friendship Paradox and Degree Correlation on Information Diffusion under Monophilic Contagion}
%

\begin{figure*}[]
	\centering
	\begin{subfigure}[!h]{0.45\textwidth}
		\centering
		\includegraphics[height = 2.75in, width = 2.5in]{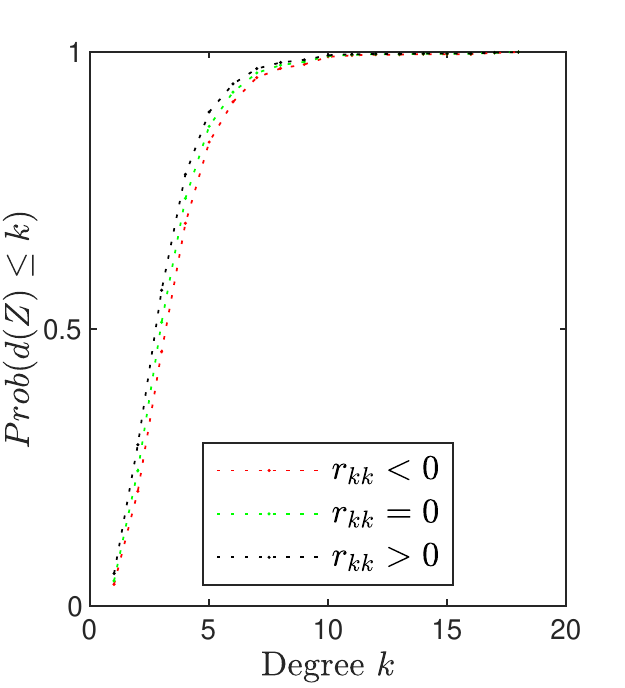}
		\caption{CDFs of the of the degree $d(Z)$  of a random friend $Z$ of a random node for three networks with same degree distribution but different assortativity $r_{kk}$ values. Note that the CDFs are point-wise increasing with $r_{kk}$ showing that $\mathbb{E}\{d(Z)\}$ decreases with $r_{kk}$.}
		\label{subfig:cdf_degree_z_with_rkk}
	\end{subfigure}\hfill
	\begin{subfigure}[!h]{0.45\textwidth}
		\centering
		\includegraphics[height = 2.85in, width = 2.25in]{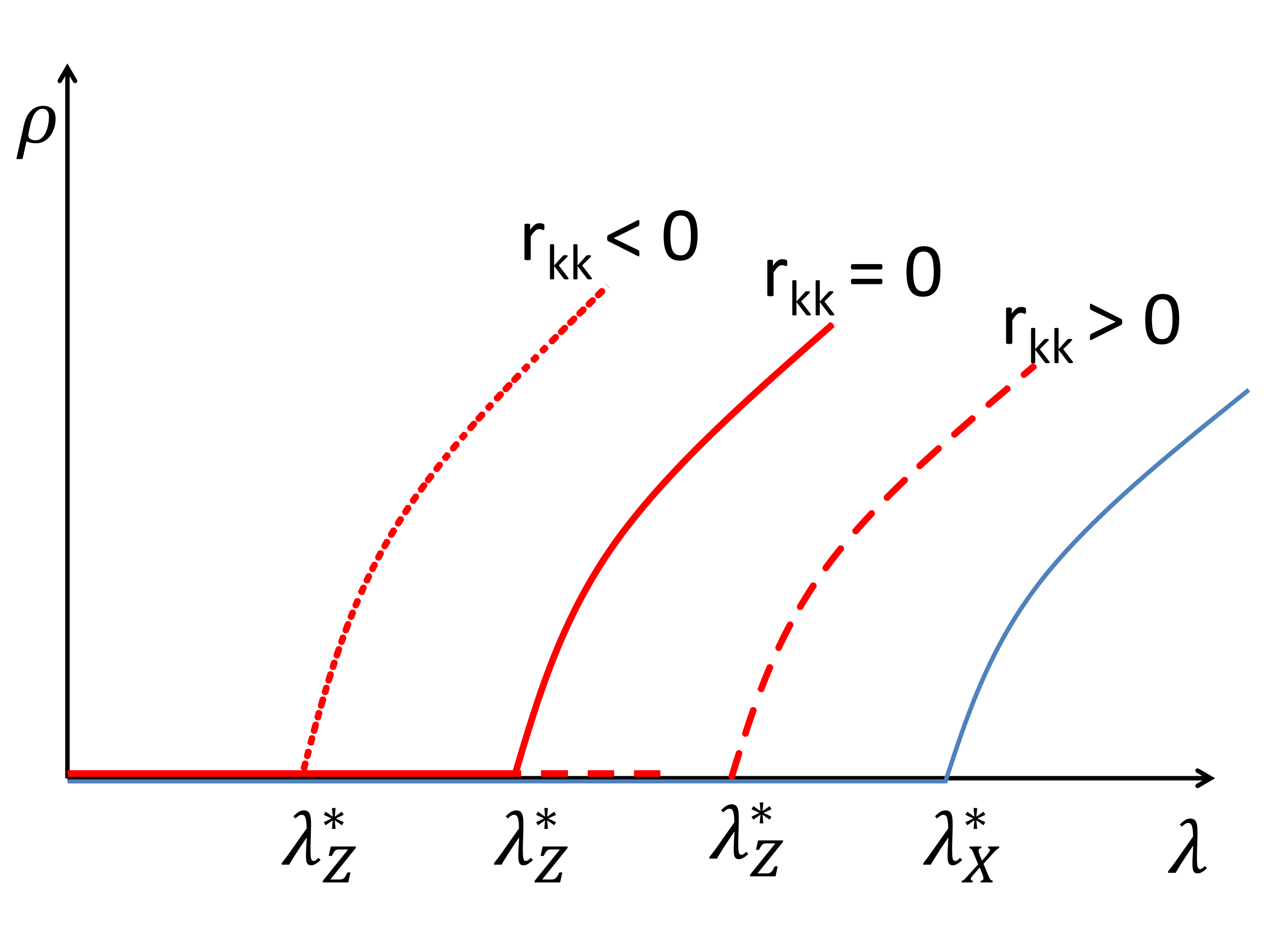}
		\caption{Variation of the stationary fraction $\rho$ of infected nodes with the effective spreading rate $\lambda$ for the case 1 (blue) and case 2 (red), illustrating the ordering of the critical thresholds of cases 1,2 and the effect of assortativity.}
		\label{subfig:stationary_infected_fraction}
	\end{subfigure}\hfill
	\caption{Comparison of non-monophilic and monophilic contagion rules and the effect of assortativity on the critical thresholds of the monophilic contagion.}
	\label{fig:assortativity_effect}
\end{figure*}

Theorem \ref{thm:observe_XZ} showed that the critical thresholds of the mean-filed dynamics equation (\ref{eq:MFD_approximation_X}) for the two rules (non-monophilic and monophilic contagion) are different. Following is an immediate corollary of Theorem \ref{thm:observe_XZ} which gives the ordering of these critical thresholds using the friendship paradox stated in Theorem \ref{th:friendship_paradox}. 
\begin{corollary}
	\label{cor:ordering_crtical_thresholds}
	The critical thresholds $\lambda^*_X, \lambda^*_Z$ in (\ref{eq:diff_threshold_X}), (\ref{eq:diff_threshold_Z}) for the cases of non-monophilic (case 1) and monophilic (case 2) contagion rules satisfy
	\begin{equation}
	\lambda^*_Z \leq \lambda^*_X.
	\end{equation}
\end{corollary}

Corollary \ref{cor:ordering_crtical_thresholds} shows that in the case of information diffusion under monophilic contagion rule, it is easier (smaller effective spreading rate) for the information to spread to a positive fraction of the agents as a result of the friendship paradox. Hence, observing random friends of random neighbors makes it easier for the information to spread instead of dying away (in unbiased-degree networks). This shows how friendship paradox can affect information diffusion over a network under monophilic contagion. 

\begin{remark}
	If we interpret an individual's second-hop connections as weak-ties, then Theorem \ref{thm:observe_XZ} and Corollary \ref{cor:ordering_crtical_thresholds} can be interpreted as results showing the importance of weak-ties in information diffusion (in the context of a SIS model and an unbiased-degree network). See the seminal works in \cite{rapoport1953spread, granovetter1973strength} for the definitions and importance of weak-ties in the sociology context. 
\end{remark}

The ordering $\lambda^*_Z \leq \lambda^*_X$ of the critical thresholds in Corollary \ref{cor:ordering_crtical_thresholds} holds irrespective of any other network property. However, the magnitude of the difference of the critical thresholds $\lambda^*_X - \lambda^*_Z$ depends on the neighbor-degree correlation (assortativity) coefficient defined as,
\begin{equation}
\label{eq:deg_deg_corr}
r_{kk} = \frac{1}{\sigma_q^2}\sum_	{k,k'}kk'\Big(e(k,k')  - q(k)q(k')\Big)
\end{equation} using the notation defined in Footnote \ref{fn:jdd}. To intuitively understand this, consider a star graph that has a negative assortativity coefficient (as all low degree nodes are connected to the only high degree node). Therefore, a randomly chosen node $X$ from the star graph has a much smaller expected degree $\mathbb{E}\{d(X)\}$ than the expected degree $\mathbb{E}\{d(Z)\}$ of a random friend $Z$ of the random node $X$ compared to the case where the network has a positive assortativity coefficient. This phenomenon is further illustrated in Fig. \ref{subfig:cdf_degree_z_with_rkk} using three networks with the same degree distribution but different assortativity coefficients obtained using Newman's edge rewiring procedure  \cite{newman2002assortative}.

Consider the stationary fraction of the infected nodes \begin{equation}
\label{eq:stationary_infected_fraction}
\rho = \sum_{k}P(k)x(k)
\end{equation} where $P(k)$ is the degree distribution and $x(k), k = 1,\dots, D$ are the stationary states of the mean-field dynamics in (\ref{eq:MFD_approximation_X}). Fig. \ref{subfig:stationary_infected_fraction} illustrates how the stationary fraction of the infected nodes varies with the effective spreading rate $\lambda$ for case 1 and 2, showing the difference between the two cases and the effect of assortativity.

\section{Collective Dynamics of SIS-Model and Reactive Networks}
\label{sec:active_networks}

So far in Sec. \ref{sec:preliminaries} and Sec. \ref{sec:effects_FP_SISModel}, the underlying social network on which the information diffusion takes place was treated as a deterministic graph and, the mean-field dynamics equation (\ref{eq:MFD_approximation_X}) was used to approximate the SIS-model. In contrast, this section explores the more general case where the underlying social network also randomly evolve at each time step $n$ (of the SIS-model) in a manner that depends on the population state $\bar{x}_n$. Our aim is to obtain a tractable model that represent the collective dynamics of the SIS-model and the evolving graph process. As explained in Sec. \ref{subsec:main_results_organization} with examples, the motivation for this problem comes from the real world networks that evolves depending on the state of information diffusion on them. In order to state the main result, we first define a reactive network and state our assumptions.

\begin{definition}[Reactive Network]
	\label{defn:reactive_network}
	A reactive network is a Markovian graph process $\{G_n\}_{n\geq 0}$ with a state space $\mathcal{G} =  \{\mathcal{G}_1,\dots,\mathcal{G}_N\}$ consisting of $N$ graphs and transition probabilities parameterized by the population state $\bar{x}_n$ i.e. $G_{n+1} \sim P_{\bar{x}_n}(\,\cdot\,|G_n)$.
\end{definition}

In Definition \ref{defn:reactive_network}, the parameterization of the transition probabilities by the population state $\bar{x}_{n}$ represents the (functional) dependency of the graph process on the current state of the SIS information diffusion process. The name reactive network is derived from this functional dependency of the graph evolution on the population state. We make the following two assumptions on the reactive graph process (Definition \ref{defn:reactive_network}). 

\begin{assumption}
	\label{assumption:reactiveNW_deg_dist}
Each graph $\mathcal{G}_i \in \mathcal{G},\, i = 1,\dots, N$ has the same degree distribution $P(k)$ but different conditional degree distributions $P_{\mathcal{G}_1}(k|k'),\dots, P_{\mathcal{G}_N}(k|k')$.
\end{assumption}

\begin{assumption}
	\label{assumption:reactiveNW_MC}
	The transition probability matrix $P_{\bar{x}_n}$ of the reactive network $\{G_n\}_{n\geq 0}$ (Definition \ref{defn:reactive_network}) is irreducible and positive recurrent with a unique stationary distribution $\pi_{\bar{x}_n}$ for all values of the population state $\bar{x}_n$.
\end{assumption}


The first assumption imposes the constraint that each graph in the state space has the same degree distribution. Under this assumption, the number of nodes $M(k)$ with degree $k$ will remain the same at each time instant $n$ and hence, the new population state at each time instant can still be expressed as the old population state plus an update term as in Theorem \ref{th:MFD}. The second assumption is standard in Markov chains and it ensures the convergence to a unique stationary distribution.

In this context, the main result of this section is the following (proof given in \cite{nettasinghe2018contagions}).
\begin{theorem}[Collective Dynamics of SIS-model and Reactive Network]
	\label{th:active_network}
	Consider a reactive network $\{G_n\}_{n\geq 0}$ (Definition \ref{defn:reactive_network}) with state space $\mathcal{G}$ and transition probabilities $P_{\bar{x}_n}(\,\cdot\,|G_n)$ (parameterized by the population state $\bar{x}_n$) satisfying the Assumptions \ref{assumption:reactiveNW_deg_dist} and \ref{assumption:reactiveNW_MC}. Let the $k^{th}$ element of the vector $H(x_n, G_n)$ be
	\begin{align}
	H_k(x_n,G_n) &= (1- x_n(k)) \frac{\nu k \theta_n^Z}{D} - x_n(k)\delta \quad \textit{where},\\
	\theta_n^Z &= \sum_{k}\bigg(\sum_{k'}P(k')P_{G_n}(k|k')\bigg) x_n(k).
	\end{align} Further, assume that $H(x,\mathcal{G}_i)$ is Lipschitz continuous in $x$ for all $\mathcal{G}_i\in \mathcal{G}$. Then, the sequence of the population state vectors $\{\bar{x}_n\}_{n\geq 0}$ generated by the SIS model under monophilic contagion over the reactive network converges weakly to the trajectory of the deterministic differential equation
	\begin{align}
	\label{eq:ODE_active_network}
	\frac{dx}{dt} &= \mathbb{E}_{G\sim \pi_x}\{H(x,G)\} \quad\text{(ODE)}\\
	\label{eq:algebraic_constraint}
	P'_x\pi_x &= \pi_x. \quad\text{(algebraic constraint)}
	\end{align}
\end{theorem}

 Theorem \ref{th:active_network} shows that the dynamics of the population state of the SIS model for information diffusion over a reactive network (Definition \ref{defn:reactive_network}) can be approximated by an ODE with an algebraic constraint. From a statistical modeling perspective, Theorem \ref{th:active_network} provides a useful means of approximating the complex dynamics of two interdependent stochastic processes (information diffusion process and the stochastic graph process) by an ODE (\ref{eq:ODE_active_network}) whose trajectory $x(t)$ at each time instant $t>0 $ is constrained by the algebraic condition (\ref{eq:algebraic_constraint}). Further, having an algebraic constraint restricts the number of possible sample paths of the population state vector $\{\bar{x}_n\}_{n\geq 0}$. Hence, from a statistical inference/filtering perspective, this makes estimation/prediction of the population state easier. For example, the algebraic condition can be used in Bayesian filtering algorithms (such as the one discussed in Sec. \ref{sec: NL_Filtering}) for the population state  to obtain more accurate results.

\section{Friendship Paradox based Polling for Networks}
\label{sec:FP_based_polling}

In Sec. \ref{sec:effects_FP_SISModel} and Sec. \ref{sec:active_networks}, the effects of the friendship paradox on SIS-model and the effects of state dependent network evolutions were discussed. In contrast, this section deals with polling: estimating the fraction of infected (state denoted by $1$) individuals 
\begin{equation}
\truevalue_n = \sum_{k}P(k)x_n(k)
\end{equation}
at a given time instant $n$, using the responses (to some query) of  $b$ sampled individuals from the network. It is assumed that the information diffusion is slow and the states of nodes remain unchanged during the estimation task. In other words, we assume that the information diffusion takes place on a slower time scale compared to the time it take to estimate $\truevalue_n$. 
 
 \bigskip
 \noindent
 {\bf Notation:} Since we consider the case of estimating the fraction of infected nodes at a given time instant $n$, we omit the subscript denoting time and use $\truevalue$ and $s^{(\cdot)}$ to denote the infected fraction of nodes and state of nodes respectively (at the given time instant  $n$) in this section. 
 \bigskip

\noindent
{\bf Motivation and Related Work:} Recall that in intent polling\footnote{This method is called intent polling because, in the case of predicting the outcome of an election, this is equivalent to asking the voting intention of sampled individuals i.e. asking ``Who are you going to vote for in the upcoming election?'') \cite{rothschild2011}.}, a set $S$ of nodes are obtained by uniform sampling with replacement and then, the average of the labels $s^{(u)}$ of nodes $u\in S$
\begin{equation}
\label{eq:intent_polling_estimate}
\estIP = \frac{\sum_{u\in S}s^{(u)}}{\vert S \vert},
\end{equation} is used as the estimate (called intent polling estimate henceforth) of the fraction $\truevalue$ of infected individuals.  The main limitation of intent polling is that the sample size needed to achieve an $\epsilon$- additive error is $O(\frac{1}{\epsilon^2})$\cite{dasgupta2012}. The algorithms presented in this section are motivated by two recently proposed methods, namely ``expectation polling" \cite{rothschild2011} and ``social sampling" \cite{dasgupta2012}, that attempt to overcome this limitation in intent polling. Firstly, in expectation polling \cite{rothschild2011}, each sampled individual is asked to provide an estimate about the state held by the majority of the individuals in the network (e.g. asking ``What do you think the state of the majority is?"). Then, each sampled individual will look at his/her neighbors and provide an answer (1 or 0) based on the state held by the majority of them. This method is more efficient (in terms of sample size) compared to the intent polling method since each sample now provides the putative response of a neighborhood\footnote{Intent polling and expectation polling have been considered intensively in literature, mostly in the context of forecasting elections and, it is generally accepted that expectation polling is more efficient compared to intent polling \cite{graefe2015accuracy,graefe2014accuracy,murr2011wisdom,murr2015wisdom,manski2004measuring}.}$^{,}$\footnote{\cite{krishnamurthy2014interactive,krishnamurthy2016} discuss how expectation polling can give rise to misinformation propagation in social learning and, propose Bayesian filtering methods to eliminate the misinformation propagation.}. Secondly, in social sampling \cite{dasgupta2012}, the response of each sampled individual is a function of the states, degrees and the sampling probabilities of his/her neighbors. \cite{dasgupta2012} provides several unbiased estimators for the fraction $\truevalue$ using this method and, establishes bounds for their variances.  The main limitation of social sampling method (compared to friendship paradox based algorithms in Sec. \ref{sec:FP_based_polling}) is that it requires the sampled individuals to know a significant amount of information about their neighbors (apart from just their labels), the graph and the sampling process (employed by the pollster). Therefore, a practical implementation of social sampling in the setting of estimating the fraction of infected individuals at a given time instant is not practically feasible. These facts motivate the polling method called {neighborhood expectation polling (NEP)} \cite{nettasinghe2018your} which we present next.

\bigskip
In NEP, a set $S \subset V$ of individuals from the social network $G = (V,E)$ are selected and asked, 
\begin{center}
	\textit{``What is your estimate of the fraction of people with label 1?"}.
\end{center}
When trying to estimate an unknown quantity about the world, any individual naturally looks at his/her neighbors.  Therefore, each sampled individual $s \in S$ would provide the fraction of their neighbors $\mathcal{N}(s)$, with label $1$. In other words, the response of the individual $s\in S$ for the NEP query would be, 
\begin{align}
q(s) &= \frac{\vert\{u \in \mathcal{N}(s): s^{(u)} = 1 \}\vert}{\vert \mathcal{N}(s)\vert}. \label{eq:response_to_query}
\end{align}
Then, the average of all the responses $\frac{\sum_{s \in S}q(s)}{\vert S \vert}$ is used as the NEP estimate of the fraction $\truevalue$.

\bigskip
\noindent
{\bf Why call it NEP?} The term neighborhood expectation polling is derived, from the fact that the response $q(v)$ of each sampled individual $v\in S$ is the expected label value among her neighbors i.e. ${q(v) = \mathbb{E}\{s^{(U)}\}}$ where, $U$ is a random neighbor of the sampled individual $v\in S$.

\begin{figure}[]
		\centering
			\label{fig:examples_error_NEP}
	\begin{subfigure}[!h]{\textwidth}
		\centering
		\includegraphics[width=2.3in]{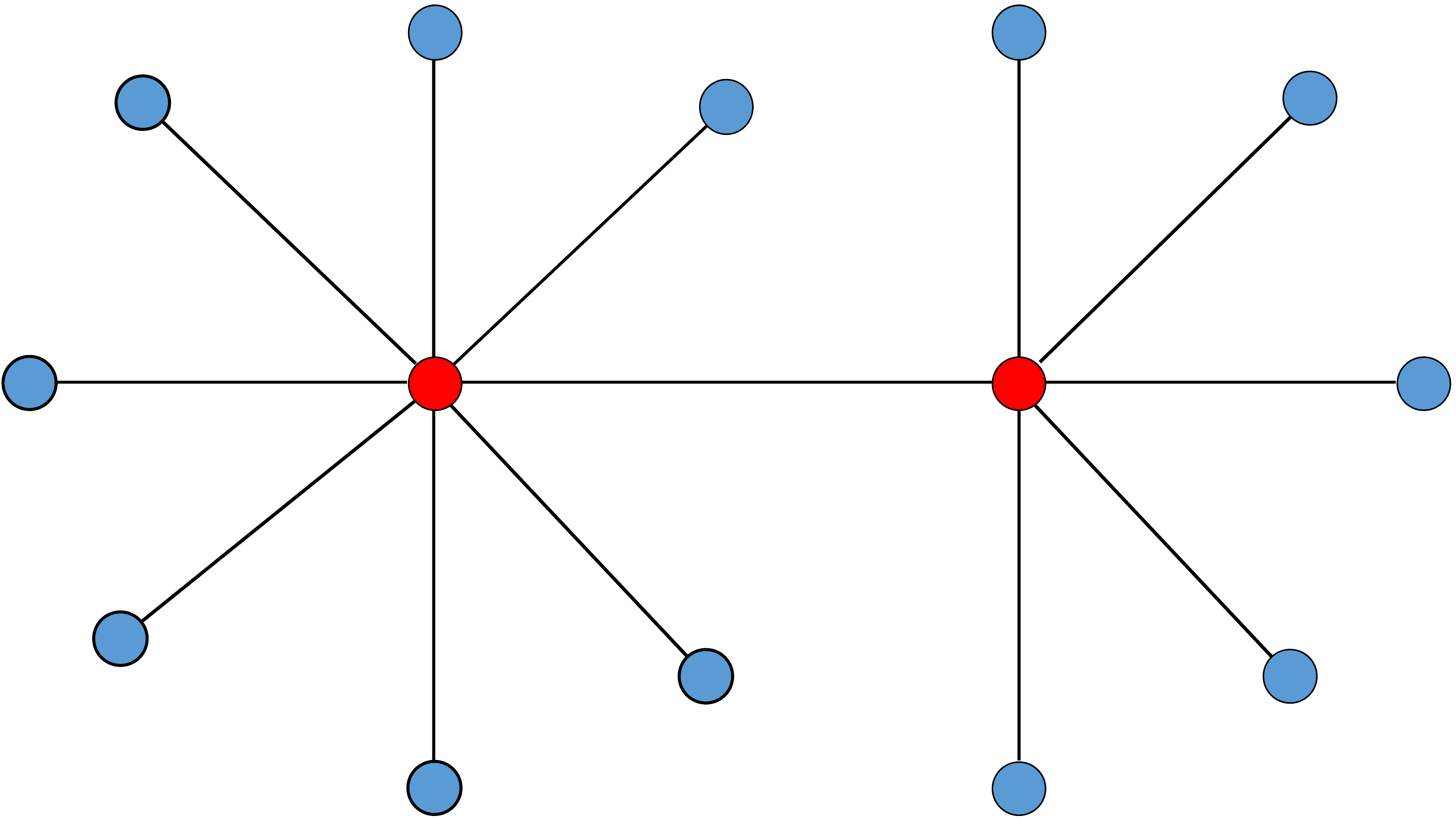}
		\caption{Network $G_1$: labels are highly correlated with the degrees of nodes }
		\label{subfig:NEP_larger_bias}
	\end{subfigure}\newline
	\begin{subfigure}[!h]{\textwidth}
		\centering
		\includegraphics[width=2.3in]{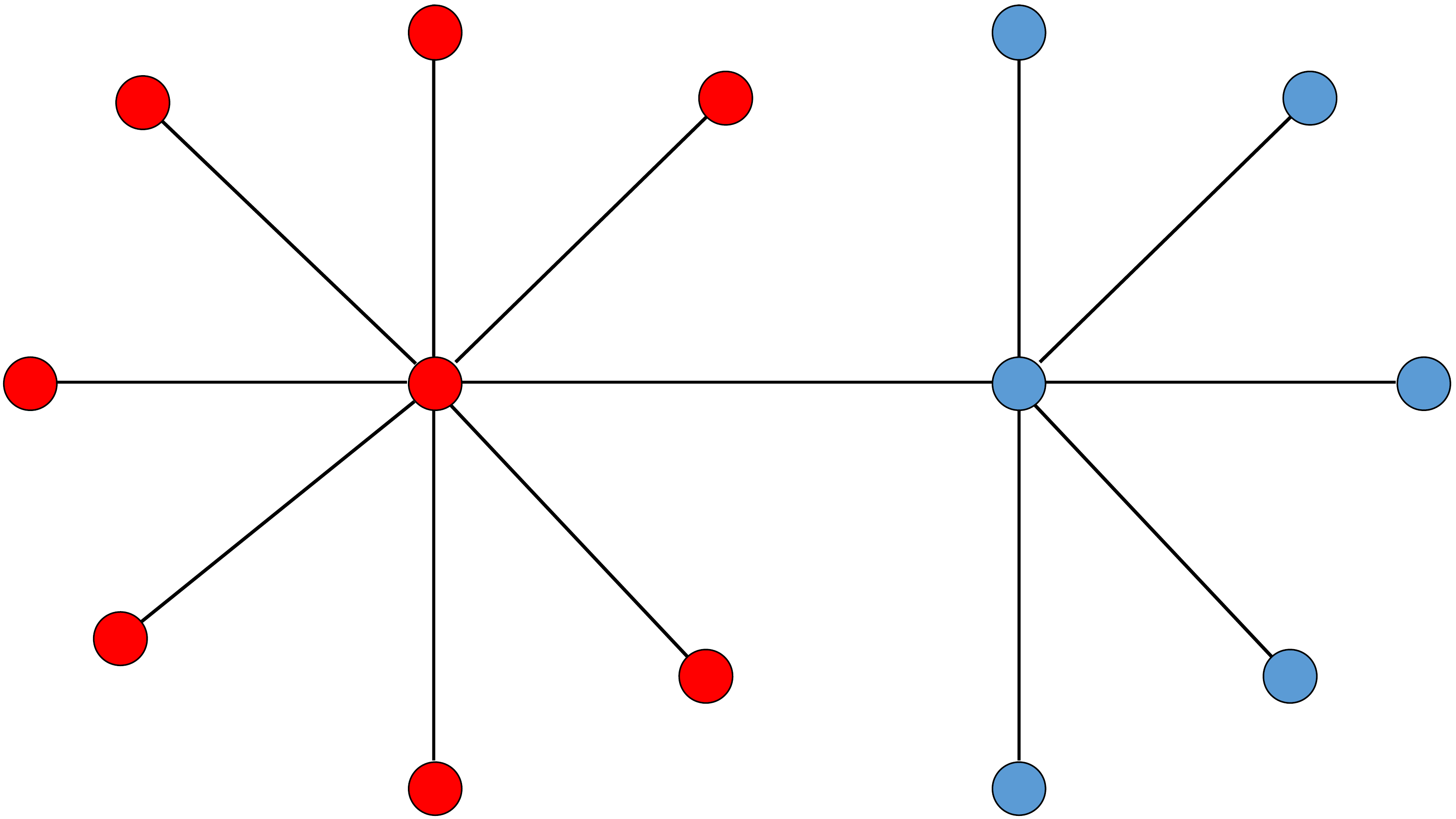}
		\caption{Network $G_2$: nodes with the same label are clustered (depicting \textit{Homophily})}
		\label{subfig:NEP_large_variance}
	\end{subfigure}\newline
	\begin{subfigure}[!h]{\textwidth}
		\centering
		\includegraphics[width=2.3in]{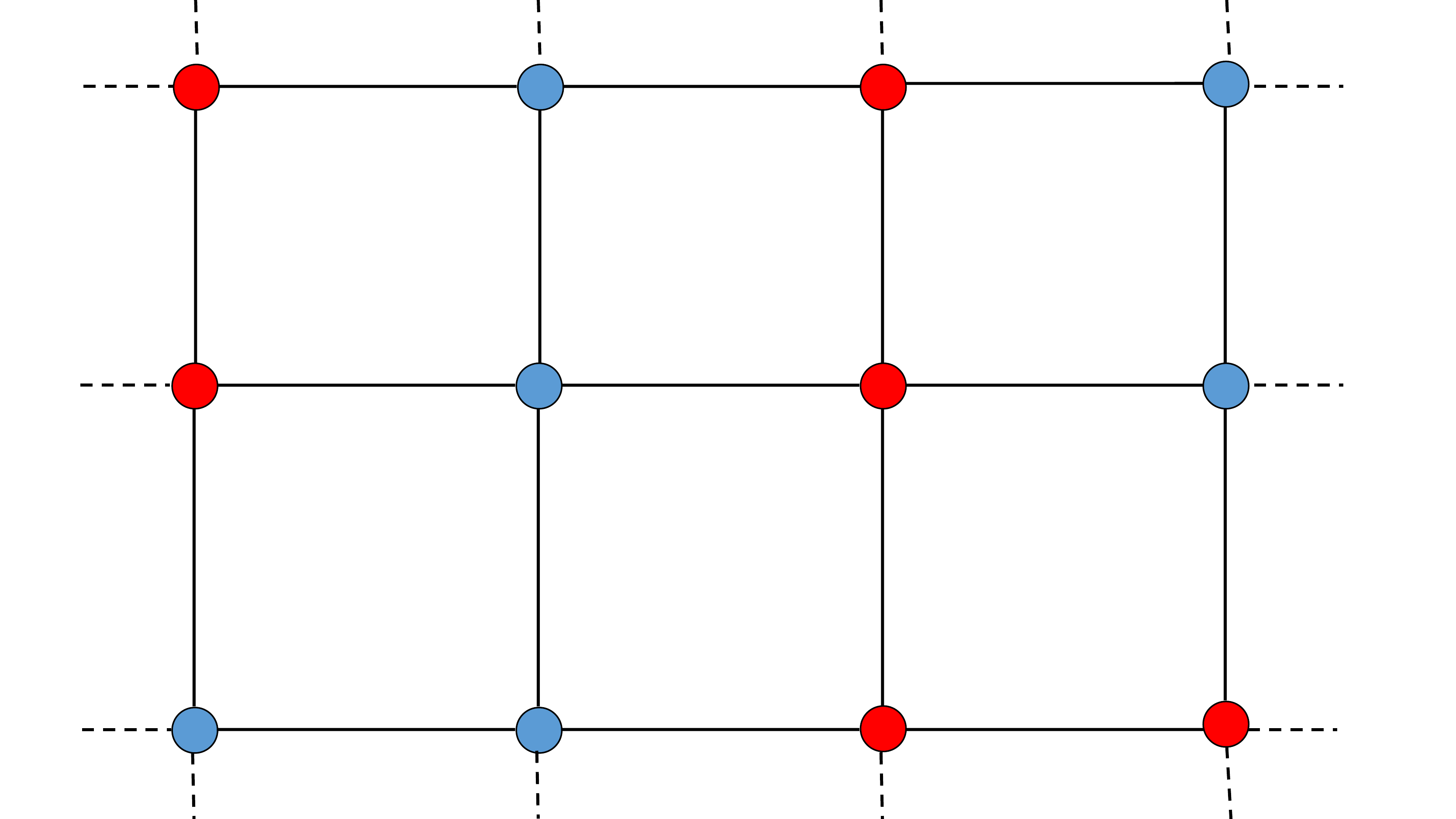}
		\caption{Network $G_3$: a large regular graph with uniformly at random assigned labels}
		\label{subfig:NEP_unbiased}
	\end{subfigure}
	
		\caption{Consider the case of uniformly sampling nodes and obtaining responses $q(s)$ of sampled nodes $s\in S$ about the fraction of red (i.e. label 1) nodes in the network. In graph $G_1$ of Fig. \ref{subfig:NEP_larger_bias}, most nodes have their only neighbor to be of color red even though most of the nodes in the network are of color blue. Hence, uniformly sampling nodes for NEP in this case would result in a highly biased estimate. In graph $G_2$ of Fig. \ref{subfig:NEP_large_variance}, approximately half the nodes have only a red neighbor and, rest of the nodes have only a blue neighbor. Hence, uniformly sampling nodes for NEP in this case would result in an estimate with a large variance. In graph $G_3$ of Fig. \ref{subfig:NEP_unbiased}, average of the NEP responses $q(v)$ of nodes is approximately equal to the fraction of nodes with red labels. Further, $q(v)$ does not vary largely among nodes. Hence, uniformly sampling nodes for NEP in this case would result in an accurate estimate.  Similar examples can also be found in \cite{dasgupta2012}. The figure highlights the importance of exploiting network structure and node labels when sampling nodes for NEP.}

\end{figure}

\bigskip
\noindent
{\bf Why (not) use NEP?} NEP is substantially different  to classical intent polling where, each sampled individual is asked \textit{``What is your label?"}. In intent polling, the response of each sampled individual $v\in S$ is his/her label $s^{(v)}$. In contrast, in NEP, the response $q(v)$ of each sampled individual $v\in S$ is a function of his/her neighborhood (defined by the underlying graph $G$) as well as the labels of his/her neighbors. Therefore, depending on  the graph $G$, function $s^{(\cdot)}$ and the method of obtaining the samples $S$, NEP might produce either,
\begin{enumerate}[I]
	\item an estimate with a larger MSE compared to intent polling (e.g. networks in Fig. \ref{subfig:NEP_larger_bias} and Fig. \ref{subfig:NEP_large_variance} shows when uniform sampling of individuals for NEP might not work), or,
	
	\item an estimate with a smaller MSE compared to intent polling (e.g. network in Fig. \ref{subfig:NEP_unbiased} shows when uniform sampling of individuals for NEP might work)
\end{enumerate}
These two possible outcomes highlight the importance of using the available information about the graph $G$ and the function $s^{(\cdot)}$ (which represent the states at the time of the estimation), when selecting the set $S$ of individuals in NEP. This lead us to the friendship paradox based NEP algorithms. 

\subsection{NEP Algorithms Based on Friendship Paradox}
\label{subsec:NEP_with_FP}
In this subsection, we consider randomized methods for selecting individuals for NEP based on the concept of friendship paradox explained in Sec. \ref{subsec:friendship_paradox}.

\subsubsection{Case 1 - Sampling friends using random walks}
\label{subsubsec:RW_method}

In this section, we consider the case where the graph $G = (V,E)$ is not known initially, but sequential exploration of the graph is possible using multiple random walks over the nodes of the graph. 

\bigskip
{\noindent {\bf A motivating example} for case 1 is a massive online social network where the fraction of user profiles indicating infection needs to be estimated (e.g. profiles mentioning symptoms of a disease). Web-crawling (using random walks) approaches are widely used to obtain samples from such  massive online social networks without requiring the global knowledge of the full network graph \cite{leskovec2006,gjoka2010,ribeiro2010estimating, gjoka2011practical,mislove2007measurement}. 
	
	\begin{algorithm}
		\caption{NEP with Random Walk Based Sampling}
		\label{alg:RW_Sampling}
		\DontPrintSemicolon 
		\KwIn{$\samplingbudget$ number of samples $\{v_1, v_2,\dots, v_b\} \subset V$.}
		\KwOut{Estimate $ \estRW$ of the of the fraction $\truevalue$ of nodes with label $1$.}
		
		\vspace{0.3cm}
		
		\begin{enumerate}
			\item Initialize $\samplingbudget$ random walks on the social network starting from $v_1, v_2,\dots, v_b$.

			\item Run each random walk for a $N$ steps and then collect sample $S = \{s_i,\dots, s_b\}$ where, $s_i \in V$ is collected from i\textsuperscript{th} random walk.

			\item Query each $s \in S$ to obtain $q(s)$ and, compute the estimate
			\begin{equation}
			\estRW =  \frac{\sum_{s \in S} q(s)}{b} \nonumber
			\end{equation} of the fraction $\truevalue$ of nodes with label $1$.
			
		\end{enumerate}
		
	\end{algorithm}
Algorithm \ref{alg:RW_Sampling} was proposed in \cite{nettasinghe2018your} for estimating the fraction $\truevalue$ in case 1. The intuition behind Algorithm \ref{alg:RW_Sampling} stems from the fact that the stationary distribution of a random walk on an undirected graph is uniform over the set of neighbors \cite{aldous2002reversible}. Therefore, Algorithm \ref{alg:RW_Sampling} obtains a set $S$ of $\samplingbudget$ neighbors independently (for sufficiently large $N$) from the graph $G = (V,E)$ in step 2. Then, the response $q(s)$ of each sampled individual $s\in S$ for the NEP query is used to compute the estimate $\estRW$ in step 3. According to the friendship paradox (Theorem \ref{th:friendship_paradox}), using uniformly sampled neighbors is equivalent to using more nodes due to the fact that random neighbors have more neighbors than random nodes on average. Hence, it is intuitive that the performance of this method should have a smaller MSE compared to the method of NEP with uniformly sampled nodes and intent polling method. In Sec. \ref{subsec:analysis}, we verify this claim theoretically and explore the conditions on the state function $s^{(\cdot)}$ and the properties of the graph $G$ for the estimator $\estRW$ to be more accurate compared to the intent polling method.
	
	\subsubsection{Case 2 - Sampling a Random Friend of a Random Individual}
	\label{subsubsec:FN_method}

	Here we assume that the graph $G = (V,E)$ is not known and it is not possible to crawl the graph (using random walks). It is further assumed that a set of uniform samples $S = \{s_1,\dots, s_k\}$ from the set of nodes $V$ can be obtained and, each sampled individual $s_i \in S$ has the ability to answer the question \textit{"What is your (random) friend's estimate of the fraction of individuals with label 1?"}. 
	
	\bigskip
	{\noindent {\bf A motivating example} for case 2 is the situation where random individuals are requested to answer survey questions for an incentive. In most such cases, the pollster does not have any information about the structural connectivity of the queried individuals and, will only be able to obtain their answer for a question. 
		
		For this case, Algorithm \ref{alg:Friend_of_Node_Sampling} was proposed in \cite{nettasinghe2018your} to obtain an estimate of the fraction $\truevalue$ of individuals with label $1$.
		\begin{algorithm}
			\caption{NEP using Friends of Uniformly Sampled Nodes}
			\label{alg:Friend_of_Node_Sampling}
			\DontPrintSemicolon 
			\KwIn{$\samplingbudget$ number of uniform samples $S = \{s_1, s_2,\dots, s_b\} \subset V$.}
			\KwOut{Estimate $\estFN$ of the of the fraction $\truevalue$ of the individuals with label 1.} 
			
			\vspace{0.3cm}
			
			\begin{enumerate}
				\item Ask each $s_i \in S$ to provide $q(u_i)$ for some randomly chosen neighbor $u_i \in \mathcal{N}(s_i)$.
				
				\item Compute the estimate,
				\begin{equation}
				\estFN =  \frac{\sum_{i = 1}^{b} q(u_i)}{b}\nonumber
				\end{equation} of the fraction $\truevalue$ of the individuals with label 1. 
				
			\end{enumerate}
			
		\end{algorithm}

		In Algorithm \ref{alg:Friend_of_Node_Sampling}, each uniformly sampled individual is asked the question \textit{"What is your (random) friend's estimate of the fraction of individuals with state 1?"}. Then, each  sampled node $s_i \in S$ would provide $q(u_i)$ for some randomly chosen $u_i \in \mathcal{N}(s_i)$. The theoretical reasoning behind this method comes from Theorem \ref{th:friendship_paradox_v2_cao} in Section \ref{sec:preliminaries} which states that, a random friend of a randomly chosen individual has more friends than a randomly chosen individual on average\footnote{It should be noted that this does not follow from the original version of friendship paradox (Theorem \ref{th:friendship_paradox}) since the random friend is not a uniformly chosen neighbor from the set of all $2\vert E \vert$ neighbors. Instead, the response now comes from a random neighbor conditioned to be a friend of the sampled node. }. Therefore, intuitively this method should result in a smaller MSE compared to the method of NEP with uniformly sampled nodes and intent polling method.  
		
		\subsection{Analysis of the Estimates Obtained via Algorithms \ref{alg:RW_Sampling} and \ref{alg:Friend_of_Node_Sampling}}
		\label{subsec:analysis}
		
		Algorithm \ref{alg:RW_Sampling} and Algorithm \ref{alg:Friend_of_Node_Sampling} presented in Sec. \ref{subsec:NEP_with_FP} query random friends and random friends of random nodes (denoted by $Y, Z$ in Theorem \ref{th:friendship_paradox} and Theorem \ref{th:friendship_paradox_v2_cao}) respectively, exploiting the friendship paradox. 
		
		In this context, the aim of this subsection is to present the following results (proof can be found in \cite{nettasinghe2018your}): 
		\begin{compactenum}
			\item Theorem \ref{th:NEP_with_vs_without_FP} motivates using friendship paradox based NEP algorithms (as opposed to NEP with uniformly sampled nodes) 
			\item Theorem \ref{th:bias_var_T2} relates bias and variance of the estimate $\estRW$ obtained using Algorithm \ref{alg:RW_Sampling} to the properties of the network. Then, Corollary \ref{cor:unbiased_mse_T2_I} gives sufficient conditions for $\estRW$ to be an unbiased estimate with a smaller mean squared error (MSE) compared to intent polling method  where, MSE of an estimate $T$ of a parameter $\truevalue$ is defined as 
			\begin{align}
			\mse\{T\} &= \mathbb{E}\{(T - \truevalue)^2\}\\
			&= \bias\{T\}^2 + \var\{T\} \label{eq:MSE}
			\end{align}
			\item Theorem \ref{th:biased_mse_T2_I} motivates the use of friendship paradox based sampling methods when the sampling budget $\samplingbudget$ is small
		\end{compactenum}
		
		\begin{theorem}
			\label{th:NEP_with_vs_without_FP}
			If the label $f(v)$ of each node $v\in V$ is independently and identically distributed then,
			\begin{align}
			\mse\{\estFN\} &\leq \mse\{\estUN\}\\
			\mse\{\estRW\} &\leq \mse\{\estUN\}
			\end{align}
			where, $\mse$ denotes mean square error defined in (\ref{eq:MSE}), $\estUN$ is the NEP estimate with $\samplingbudget$ uniformly sampled nodes and, $\estRW,\estFN$ are the estimates obtained using Algorithm \ref{alg:RW_Sampling} and Algorithm \ref{alg:Friend_of_Node_Sampling} respectively. 
		\end{theorem}
		Theorem \ref{th:NEP_with_vs_without_FP} shows that friendship paradox based sampling always has a smaller mean squared error when the node labels are independently and identically distributed (iid). This motivates the use of friendship paradox based NEP methods (Algorithm \ref{alg:RW_Sampling} and Algorithm \ref{alg:Friend_of_Node_Sampling}) instead of uniform sampling based NEP. In the subsequent results, we show that the superiority of friendship paradox based NEP algorithms over the widely used intent polling method holds for conditions less stringent than the iid assumption. 
		
		Next, we formally quantify the bias $\bias(\estRW)$ and the variance $\var(\estRW)$ of the estimator $\estRW$ obtained via Algorithm \ref{alg:RW_Sampling} as the random walk length $N$ goes to infinity and then, compare it with the widely used intent polling method. 
		
		\begin{theorem}
			\label{th:bias_var_T2}
			Let $X$ be a random node and $(U,Y)$ be a random link sampled from a connected graph. Then, as $N$ tends to infinity, the bias $\bias(\estRW)$ and the variance $\var (\estRW)$ of the estimate $\estRW$, obtained via Algorithm \ref{alg:RW_Sampling} are given by,
			\begin{align}
			\bias(\estRW) &= \mathbb{E} \{s^{(Y)}\} - \mathbb{E} \{s^{(X)}\} \label{eq:bias_T2_eq1}\\
			&= \frac{\cov\{s^{(X)},d(X)\}}{\mathbb{E}\{d(X)\}}\label{eq:bias_T2_eq2}
			\end{align}
			
			\begin{equation}
			\label{eq_var_T2}
			\var\{\estRW\} = \frac{1}{b}\cov\{s^{(Y)},q(U)\}.
			\end{equation}
		\end{theorem}

		Theorem \ref{th:bias_var_T2} provides insights into the properties of the networks for which, NEP based Algorithm 2 provides a better estimate compared to the intent polling method. Eq. (\ref{eq:bias_T2_eq1}) of Theorem \ref{th:bias_var_T2} shows that, the bias of the estimate $\estRW$ is the difference between the expected label value at a random friend, $Y$ and the expected value at a random individual, $X$. Further,  (\ref{eq:bias_T2_eq2}) shows that it is proportional to the covariance between the degree $d(X)$ and the state $s^{(X)}$ of a randomly chosen node $X$.  An immediate consequence of this result is the following corollary, which gives a sufficient condition for the estimate $\estRW$ to be unbiased and, also have a smaller variance (and therefore, a smaller MSE) compared to intent polling.
		\begin{corollary}
			\label{cor:unbiased_mse_T2_I}
			If the label $s^{(X)}$ and the degree $d(X)$ are uncorrelated and the graph is connected, the following statements hold as $N$ tends to infinity: 
			\begin{enumerate}
				\item The estimate $\estRW$, obtained via Algorithm \ref{alg:RW_Sampling} is unbiased for $\truevalue$ i.e. 
				\begin{equation}
				\mathbb{E}\{\estRW\} = \truevalue
				\end{equation}
				\item  The estimate $\estRW$, obtained via Algorithm 1 is more efficient compared to intent polling estimate $I$ in (\ref{eq:intent_polling_estimate}) i.e. 
				\begin{equation}
				\mse\{\estRW\} \leq \mse\{\estIP\}
				\end{equation}	where, $\mse$ denotes mean square error defined in (\ref{eq:MSE}).
			\end{enumerate}	
		\end{corollary}
%
%
%
%
		Theorem \ref{th:bias_var_T2} also shows that the variance of the estimate $\estRW$ is the covariance of the state $s^{(Y)}$ of a random friend $Y$ and the response $q(U)$ of her random friend $U$. 
		
		The following result gives sufficient conditions for $\estRW$ to be a more efficient (in an MSE sense) estimator compared to intent polling method (even in the presence of bias) when the sampling budget $\samplingbudget=1$. 
		\begin{theorem}
			\label{th:biased_mse_T2_I}
			Assume that the graph is connected and the sampling budget $\samplingbudget = 1$. Then, as $N$ tends to infinity, the estimate $\estRW$ has a smaller MSE compared to the intent polling estimate $I$, defined in (\ref{eq:intent_polling_estimate}), if
			\begin{equation}
			\mathbb{E}\{d(X) \vert s^{(X)} =1 \} \leq \mathbb{E}\{d(X) \vert s^{(X)} =0 \} \text{ and } \truevalue \leq 0.5
			\end{equation}
			or
			\begin{equation}
			\mathbb{E}\{d(X) \vert s^{(X)}  \} \geq \mathbb{E}\{d(X) \vert s^{(X)}  =0 \} \text{ and } \truevalue \geq 0.5.
			\end{equation}
		\end{theorem} 
		Theorem \ref{th:biased_mse_T2_I} shows that, if the expected degree of an individual with state 1 is larger (smaller) compared to the expected degree of an individual with opinion 0 and, the expected state in the network is above (below) half then, MSE of the estimate $\estRW$ is smaller than intent polling estimate $I$ in (\ref{eq:intent_polling_estimate}) when the pollster can query only one individual. This helps the pollster to incorporate prior knowledge about the current state of the diffusion and the structure of the network to decide whether its suitable to use NEP based Algorithm \ref{alg:RW_Sampling} (over the intent polling method). 
		

\subsection{Numerical Examples}
\label{sec:experiments}
In this section, results (based on \cite{nettasinghe2018your}) illustrating the performance of Algorithms \ref{alg:RW_Sampling} and \ref{alg:Friend_of_Node_Sampling} are provided. The aim of these experimental results is to evaluate the dependence of the accuracy (MSE) of the estimate of $\truevalue$ on the following three properties related to the network and the state of the information diffusion:
\begin{enumerate}
	\item {\bf Degree distribution} $P(k)$ (which is the probability that a randomly chosen node has $k$ neighbors).
	\vspace{0.25cm}
	
	\item {\bf Neighbor Degree correlation (assortativity) coefficient} $r_{kk}$ defined in (\ref{eq:deg_deg_corr})

	\vspace{0.25cm}
	\item {\bf Degree-label correlation coefficient} 
	\begin{equation}
	\label{eq:deg_label_corr}
	p_{ks} = \frac{1}{\sigma_k\sigma_s}\sum_{k}k\Big(\mathbb{P}(s^{(X)} = 1,d(X) = k) - \mathbb{P}(s^{X} = 1)P(k) \Big)
	\end{equation}
	where, $\sigma_k, \sigma_s$ are the standard deviations of the degree distribution $P(k)$ and the state (label) distribution respectively and, $P(s,k)$ is the joint distribution of the states and degrees of nodes. 
\end{enumerate}
A detailed discussion about these metrics and their effects can be found in \cite{lerman2016}.

\bigskip
\cite{nettasinghe2018your}  evaluated Algorithms \ref{alg:RW_Sampling}, \ref{alg:Friend_of_Node_Sampling}  on networks obtained using two modes: configuration mode \cite{molloy1995critical} and  Erd\H{o}s-R\'{e}nyi (G(n,p)) model \cite{newman2002random}. The resulting mean squared errors for the configuration model (power-law degree distribution) are shown in Fig. \ref{fig:mse_pl_alpha_2pt1} and Fig. \ref{fig:mse_pl_alpha_2pt4} for power-law coefficient values $\alpha = 2.1$ and $\alpha = 2.4$ respectively.  Similarly, results obtained for Erd\H{o}s-R\'{e}nyi graphs (Poisson degree distribution) are shown in Fig. \ref{fig:mse_ER_davg_50}. In the case of Erd\H{o}s-R\'{e}nyi graphs, only the assortativity coefficient $r_{kk} = 0$ is considered as it cannot be changed significantly due to the homogeneity in the degree distribution (see \cite{nettasinghe2018your} for more details on the experimental procedure).

\begin{figure*}[]
	\centering
	\begin{subfigure}[!h]{0.3\textwidth}
		\centering
		\includegraphics[width=1.5in]{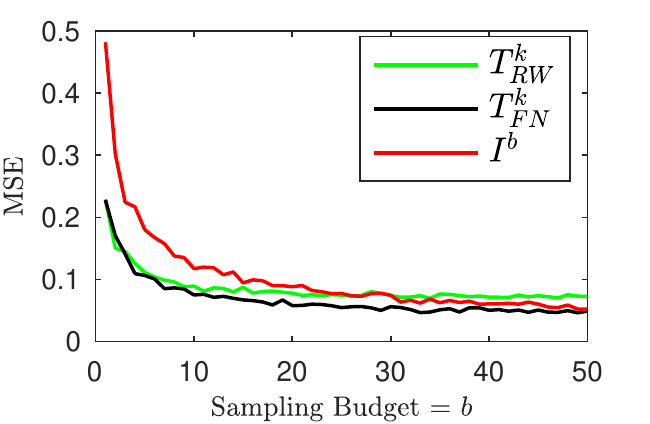}
		\caption{$r_{kk} = 0.1, p_{ks} = -0.2$}
		\label{subfig:MSE_alpha2pt1_rkk_0pt1_pkf_neg0pt2}
	\end{subfigure}\hfill
	\begin{subfigure}[!h]{0.3\textwidth}
		\centering
		\includegraphics[width=1.5in]{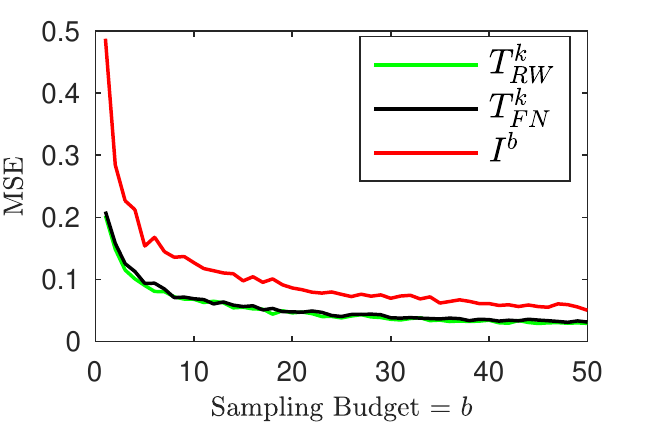}
		\caption{$r_{kk} = 0.1, p_{ks} = 0.0$}
		\label{subfig:MSE_alpha2pt1_rkk_0pt1_pkf_0}
	\end{subfigure}\hfill 
	\begin{subfigure}[!h]{0.3\textwidth}
		\centering
		\includegraphics[width=1.5in] {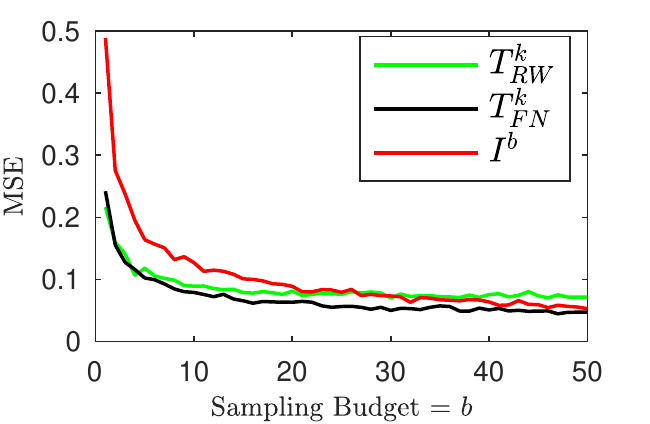}
		\caption{$r_{kk} = 0.1, p_{ks} = 0.2$}
		\label{subfig:MSE_alpha2pt1_rkk_0pt1_pkf_0pt2}
	\end{subfigure}
	\begin{subfigure}[!h]{0.3\textwidth}
		\centering
		\includegraphics[width=1.5in]{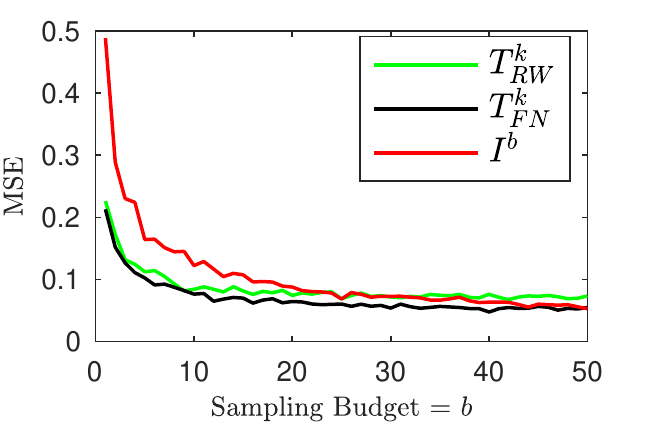}
		\caption{$r_{kk} = 0.0, p_{ks} = -0.2$}
		\label{subfig:MSE_alpha2pt1_rkk_0_pkf_neg0pt2}
	\end{subfigure}\hfill
	\begin{subfigure}[!h]{0.3\textwidth}
		\centering
		\includegraphics[width=1.5in]{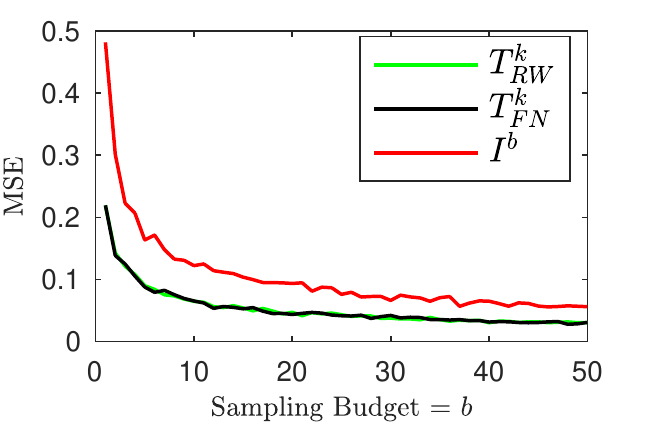}
		\caption{$r_{kk} = 0.0, p_{ks} = 0.0$}
		\label{subfig:MSE_alpha2pt1_rkk_0_pkf_0}
	\end{subfigure}\hfill 
	\begin{subfigure}[!h]{0.3\textwidth}
		\centering
		\includegraphics[width=1.5in] {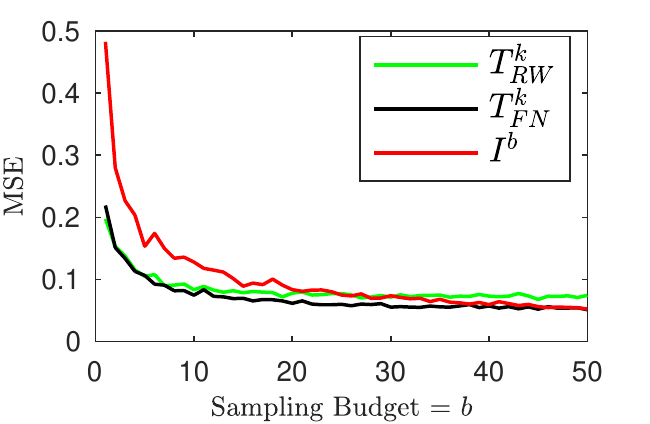}
		\caption{$r_{kk} = 0.0, p_{ks} = 0.2$}
		\label{subfig:MSE_alpha2pt1_rkk_0_pkf_0pt2}
	\end{subfigure}
	
	\begin{subfigure}[!h]{0.3\textwidth}
		\centering
		\includegraphics[width=1.5in]{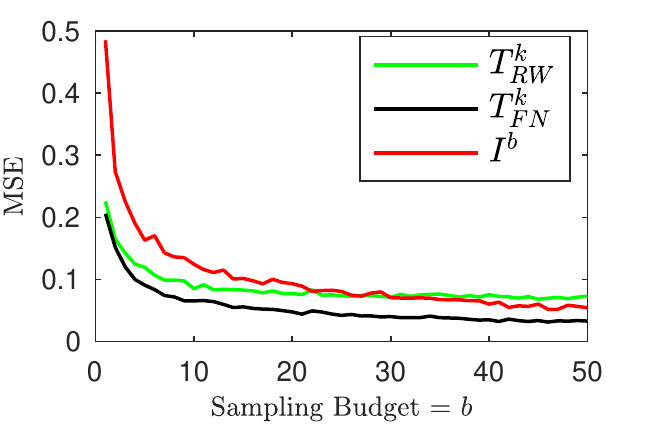}
		\caption{$r_{kk} = -0.2, p_{ks} = -0.2$}
		\label{subfig:MSE_alpha2pt1_rkk_neg0pt2_pkf_neg0pt2}
	\end{subfigure}\hfill
	\begin{subfigure}[!h]{0.3\textwidth}
		\centering
		\includegraphics[width=1.5in]{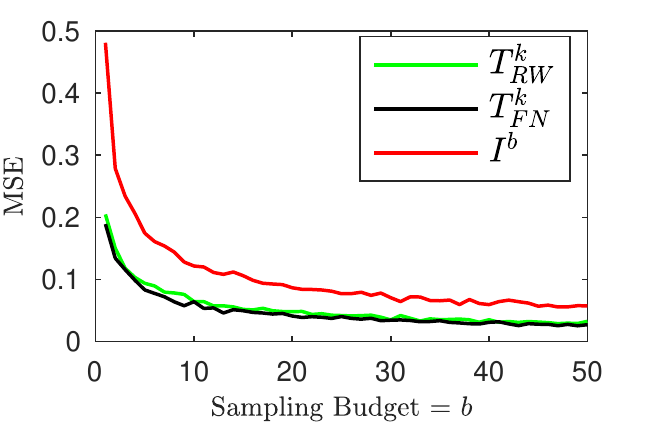}
		\caption{$r_{kk} = -0.2, p_{ks} = 0.0$}
		\label{subfig:MSE_alpha2pt1_rkk_neg0pt2_pkf_0}
	\end{subfigure}\hfill 
	\begin{subfigure}[!h]{0.3\textwidth}
		\centering
		\includegraphics[width=1.5in] {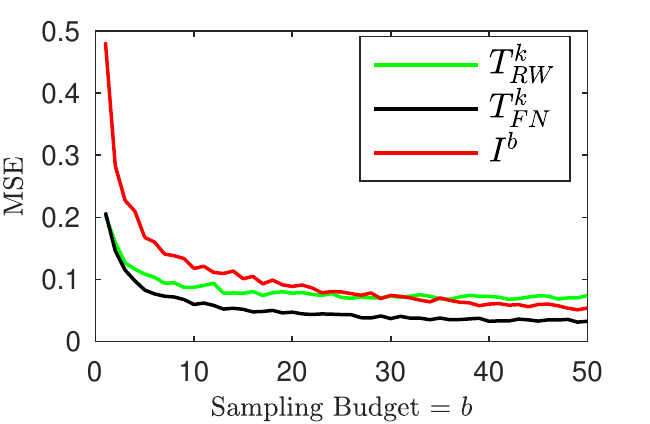}
		\caption{$r_{kk} = -0.2, p_{ks} = 0.2$}
		\label{subfig:MSE_alpha2pt1_rkk_neg0pt2_pkf_0pt2}
	\end{subfigure}
	\caption{MSE of the estimates obtained using Algorithm \ref{alg:RW_Sampling} ($\estRW$), Algorithm \ref{alg:Friend_of_Node_Sampling} ($\estFN$) and intent polling method ($\estIP$) versus the sampling budget $\samplingbudget$, for a power-law graph with parameter $\alpha = 2.1$ with different values of assortativity coefficient $r_{kk}$ and degree-label correlation coefficient $p_{ks}$. This figure shows that, for power-law networks, the proposed friendship paradox based NEP methods have smaller mean squared error compared to classical intent polling method under general conditions.}
	\label{fig:mse_pl_alpha_2pt1} 
\end{figure*}

\begin{figure*}[]
	\centering
	\begin{subfigure}[!h]{0.3\textwidth}
		\centering
		\includegraphics[width=1.5in]{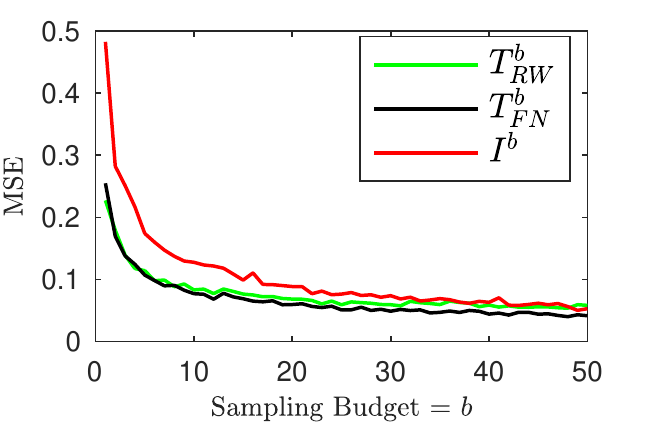}
		\caption{$r_{kk} = 0.2, p_{ks} = -0.2$}
		\label{subfig:MSE_alpha2pt4_rkk_0pt2_pkf_neg0pt2}
	\end{subfigure}\hfill
	\begin{subfigure}[!h]{0.3\textwidth}
		\centering
		\includegraphics[width=1.5in]{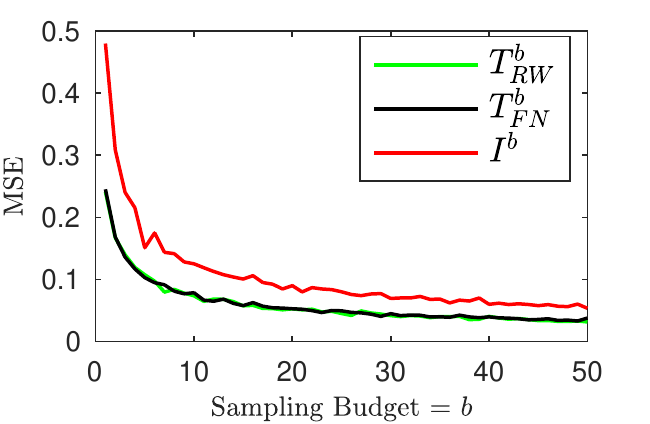}
		\caption{$r_{kk} = 0.2, p_{ks} = 0.0$}
		\label{subfig:MSE_alpha2pt4_rkk_0pt2_pkf_0}
	\end{subfigure}\hfill 
	\begin{subfigure}[!h]{0.3\textwidth}
		\centering
		\includegraphics[width=1.5in] {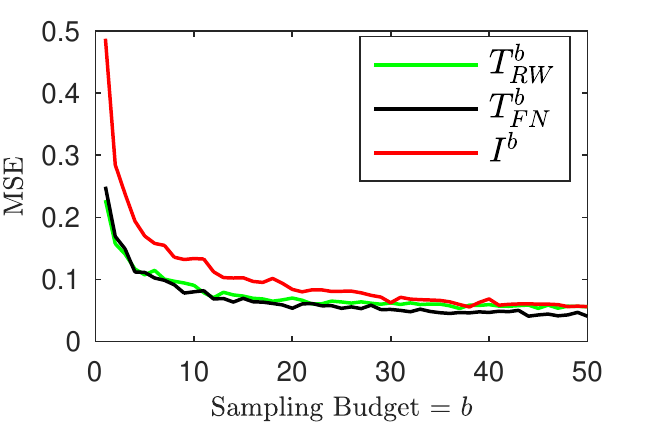}
		\caption{$r_{kk} = 0.2, p_{ks} = 0.2$}
		\label{subfig:MSE_alpha2pt4_rkk_0pt2_pkf_0pt2}
	\end{subfigure}
	
	\begin{subfigure}[!h]{0.3\textwidth}
		\centering
		\includegraphics[width=1.5in]{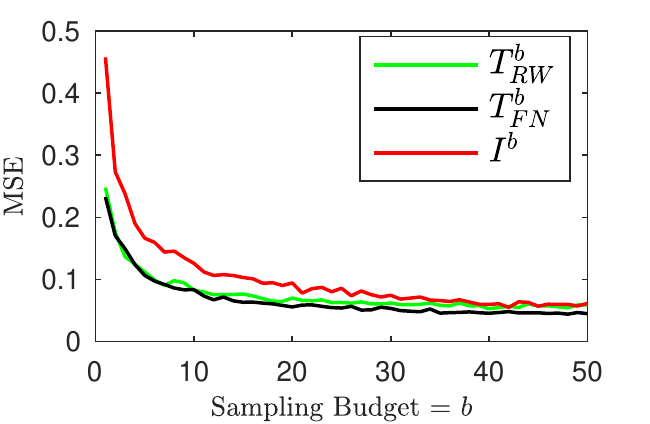}
		\caption{$r_{kk} = 0.0, p_{ks} = -0.2$}
		\label{subfig:MSE_alpha2pt4_rkk_0_pkf_neg0pt2}
	\end{subfigure}\hfill
	\begin{subfigure}[!h]{0.3\textwidth}
		\centering
		\includegraphics[width=1.5in]{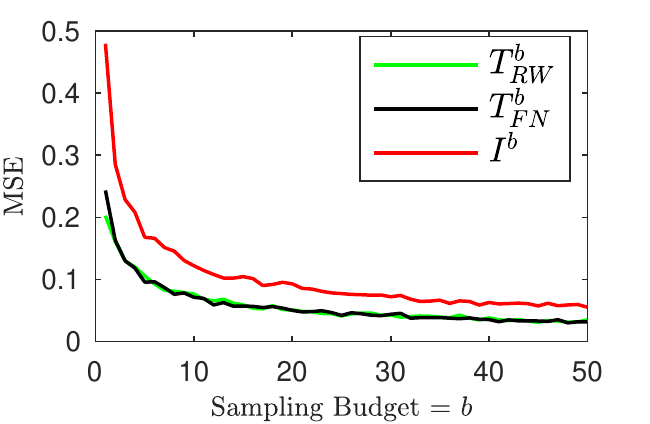}
		\caption{$r_{kk} = 0.0, p_{ks} = 0.0$}
		\label{subfig:MSE_alpha2pt4_rkk_0_pkf_0}
	\end{subfigure}\hfill 
	\begin{subfigure}[!h]{0.3\textwidth}
		\centering
		\includegraphics[width=1.5in] {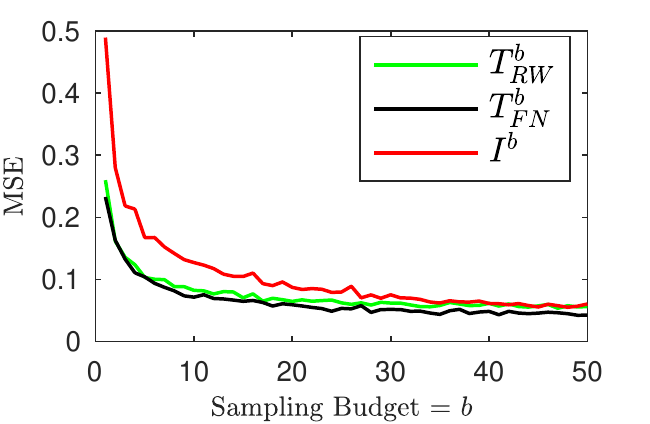}
		\caption{$r_{kk} = 0.0, p_{ks} = 0.2$}
		\label{subfig:MSE_alpha2pt4_rkk_0_pkf_0pt2}
	\end{subfigure}
	
	\begin{subfigure}[!h]{0.3\textwidth}
		\centering
		\includegraphics[width=1.5in]{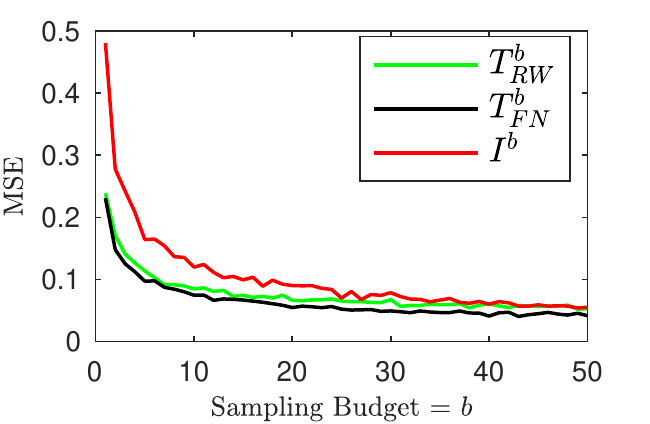}
		\caption{$r_{kk} = -0.2, p_{ks} = -0.2$}
		\label{subfig:MSE_alpha2pt4_rkk_neg0pt2_pkf_neg0pt2}
	\end{subfigure}\hfill
	\begin{subfigure}[!h]{0.3\textwidth}
		\centering
		\includegraphics[width=1.5in]{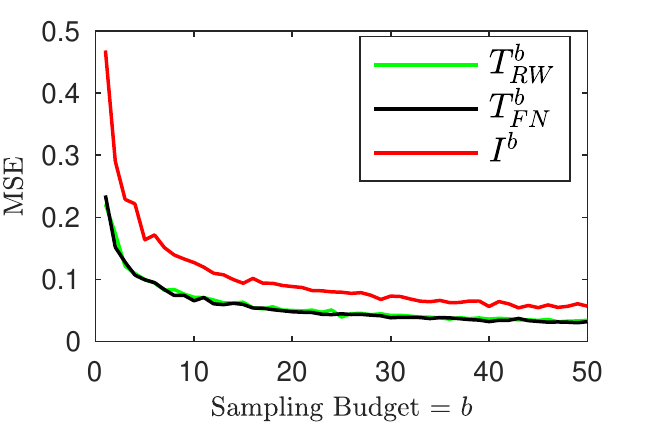}
		\caption{$r_{kk} = -0.2, p_{ks} = 0.0$}
		\label{subfig:MSE_alpha2pt4_rkk_neg0pt2_pkf_0}
	\end{subfigure}\hfill 
	\begin{subfigure}[!h]{0.3\textwidth}
		\centering
		\includegraphics[width=1.5in] {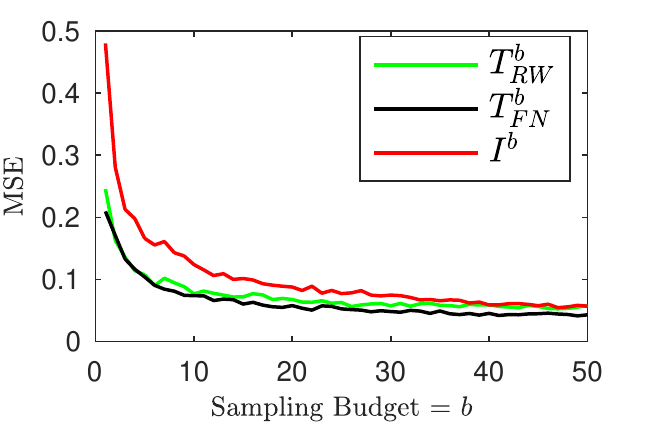}
		\caption{$r_{kk} = -0.2, p_{ks} = 0.2$}
		\label{subfig:MSE_alpha2pt4_rkk_neg0pt2_pkf_0pt2}
	\end{subfigure}
	\caption{MSE of the estimates obtained using Algorithm \ref{alg:RW_Sampling} ($\estRW$), Algorithm \ref{alg:Friend_of_Node_Sampling} ($\estFN$) and intent polling method ($\estIP$) versus the sampling budget $\samplingbudget$, for a power-law graph with parameter $\alpha = 2.4$ with different values of the assortativity coefficient $r_{kk}$ and degree-label correlation coefficient $p_{ks}$. This figure shows that, for power-law networks, the proposed friendship paradox based NEP methods have smaller mean squared error compared to classical intent polling method under general conditions.}
	\label{fig:mse_pl_alpha_2pt4} 
\end{figure*}

\begin{figure*}[]
	\centering
	\begin{subfigure}[!h]{0.3\textwidth}
		\centering
		\includegraphics[width=1.5in]{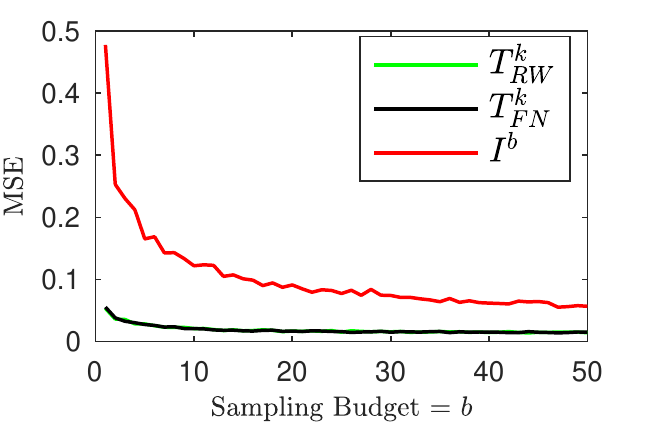}
		\caption{$r_{kk} = 0.0, p_{ks} = -0.2$}

	\end{subfigure}\hfill
	\begin{subfigure}[!h]{0.3\textwidth}
		\centering
		\includegraphics[width=1.5in]{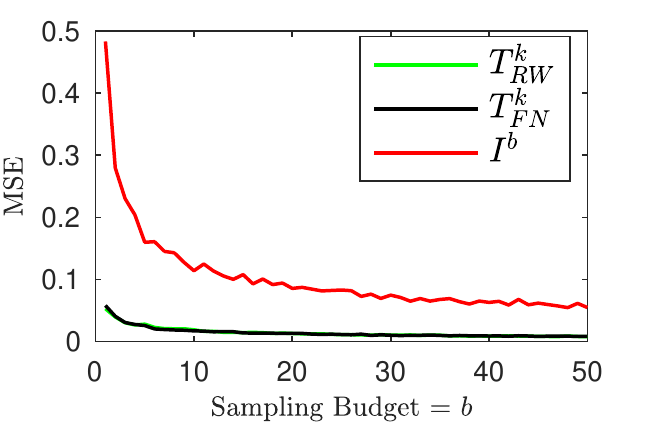}
		\caption{$r_{kk} = 0.0, p_{ks} = 0.0$}

	\end{subfigure}\hfill 
	\begin{subfigure}[!h]{0.3\textwidth}
		\centering
		\includegraphics[width=1.5in] {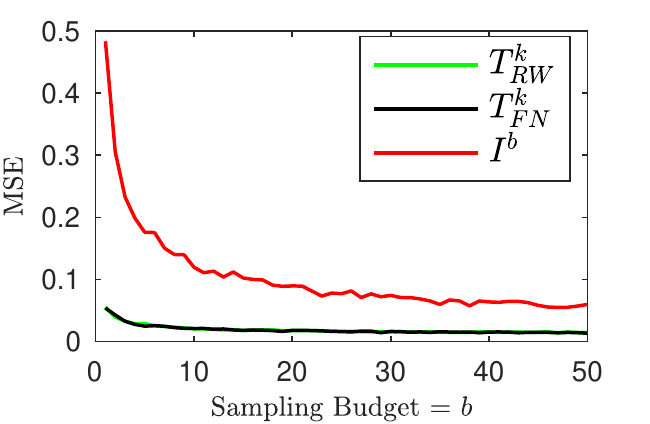}
		\caption{$r_{kk} = 0.0, p_{ks} = 0.2$}

	\end{subfigure}
	\caption{MSE of the estimates obtained using Algorithm \ref{alg:RW_Sampling} ($\estRW$), Algorithm \ref{alg:Friend_of_Node_Sampling} ($\estFN$) and intent polling method ($\estIP$) versus the sampling budget $\samplingbudget$, for a Erd\H{o}s-R\'{e}nyi  graph with parameter average degree 50 with assortativity coefficient $r_{kk} = 0$ and different values of degree-label correlation coefficient $p_{ks}$. This figure shows that, for ER graphs, the proposed friendship paradox based NEP method as well as the greedy deterministic sample selection method result in better performance compared to the intent polling method.}
	\label{fig:mse_ER_davg_50} 
\end{figure*}

\subsection{Discussion of the Results}
\label{sec:discussion}
In this subsection, the main findings of the experiments and how they relate to the theoretical results are discussed. Further, how these findings can be useful to identify the best possible algorithm (out of Algorithms \ref{alg:RW_Sampling}, Algorithm \ref{alg:Friend_of_Node_Sampling} and the alternative intent polling method) depending on the context is discussed.

\subsubsection{Power-law Graphs}
\noindent
{\bf Intent Polling vs. Friendship Paradox Based Polling: } The friendship paradox based Algorithm \ref{alg:RW_Sampling}, Algorithm \ref{alg:Friend_of_Node_Sampling} performs better for all sample sizes compared to the intent polling method when the degree-label correlation $p_{ks} = 0$ (which agrees with Theorem \ref{cor:unbiased_mse_T2_I}). Further, even when $p_{ks}$ is non-zero, the Algorithm \ref{alg:Friend_of_Node_Sampling} outperform the intent polling method for all considered samples sizes while Algorithm \ref{alg:RW_Sampling} outperforms the intent polling method for in all considered cases for small sample sizes ($\samplingbudget \leq 30$).

\bigskip
\noindent
{\bf Effect of the Heavy-Tails: } By comparing Fig.~\ref{fig:mse_pl_alpha_2pt1} with Fig.~\ref{fig:mse_pl_alpha_2pt4}, it can be seen that, when the tail of the degree distribution is heavier (smaller power-law coefficient $\alpha$), the performance of Algorithms \ref{alg:RW_Sampling}, \ref{alg:Friend_of_Node_Sampling} is better than the intent polling method for small sampling budgets. The effect of the heavy tails is more visible on the Algorithm \ref{alg:Friend_of_Node_Sampling} which performs better than intent polling method in all cases for all sample sizes. This shows that, when the sampling budget is small and the network has a heavy tail, the friendship paradox based algorithms can offer significant advantage over classical intent polling method. 

\bigskip
\noindent
{\bf Effect of the Assortativity of the Network:} 
Many different joint degree distributions $e(k,k')$ can give rise to the same neighbor degree distribution $q(k)$ (which is the marginal distribution of $e(k,k')$). This marginal distribution $q(k)$ does not capture the joint variation of the degrees a random pair of neighbors. In Algorithm \ref{alg:RW_Sampling} (which samples neighbors uniformly), the degree distribution of the samples is the neighbor degree distribution $q(k)$. Hence, the performance is not affected by the assortativity coefficient $r_{kk}$, which captures the joint variation (in terms of the joint degree distribution $e(k,k')$) of the degrees of a random pair of neighbors. This is apparent in Fig. \ref{fig:mse_pl_alpha_2pt4} where, each column (corresponding to different $r_{kk}$ values) has approximately same MSE for Algorithm \ref{alg:RW_Sampling}. However, it can be seen that, the MSE of Algorithm~\ref{alg:Friend_of_Node_Sampling} (that samples random friends $Z$ of random nodes) increases with $r_{kk}$ due to the fact that the degree $d(Z)$ of a random friend $Z$ of a random node is a function of the joint degree distribution. In order to make this point clear, Fig. \ref{fig:fosdYZ_alpha2pt4} illustrates the effect of the neighbor degree correlation $r_{kk}$ on $d(Z)$ (and the invariance of $d(Y)$ to $r_{kk}$). Hence, if it is apriori known that the network is disassortative, the Algorithm \ref{alg:Friend_of_Node_Sampling} is a more suitable choice for polling (compared to Algorithm \ref{alg:RW_Sampling}).

\bigskip
\noindent
{\bf When to use friendship paradox based NEP?} Both theoretical (Theorem \ref{th:biased_mse_T2_I}) as well as numerical results (Fig. \ref{fig:mse_pl_alpha_2pt4}, Fig. \ref{fig:mse_ER_davg_50}) show that friendship paradox based NEP methods outperform classical intent polling method by a large margin when the sampling budget is small compared to the size of the network (which is the case in many applications related to polling). Further, the absence of correlation guarantees the better performance of friendship paradox based NEP methods for any sample size (Corollary \ref{cor:unbiased_mse_T2_I}) and the presence of assortativity improves the performance of Algorithm \ref{alg:Friend_of_Node_Sampling}. These results/observations gives the pollster the ability to decide which algorithm to be deployed using the available information about the network and the sampling budget. 

\subsubsection{Erd\H{o}s-R\'{e}nyi Graphs}
Erd\H{o}s-R\'{e}nyi ($G(n, p)$) model starts with $n$ vertices and then connects each two vertices with probability $p$ resulting in an average degree of $(n-1)p$. From the Fig. \ref{fig:mse_ER_davg_50}, it can be seen that Algorithms \ref{alg:RW_Sampling} and \ref{alg:Friend_of_Node_Sampling} both yield a smaller MSE than the intent polling method for Erd\H{o}s-R\'{e}nyi models (with an average degree of 50). Further, Algorithm \ref{alg:RW_Sampling} and Algorithm \ref{alg:Friend_of_Node_Sampling} both have an equal MSE in the case of Erd\H{o}s-R\'{e}nyi Graphs. This is a result of the fact that distributions of the degree $d(Y)$ of a random neighbor $Y$ and the distribution of the degree $d(Z)$ of a random neighbor $Z$ of a random node are equal when the neighbor degree correlation is zero.

\begin{figure*}[]
	\centering
	\begin{subfigure}[!h]{0.3\textwidth}
		\centering
		\includegraphics[width=1.5in]{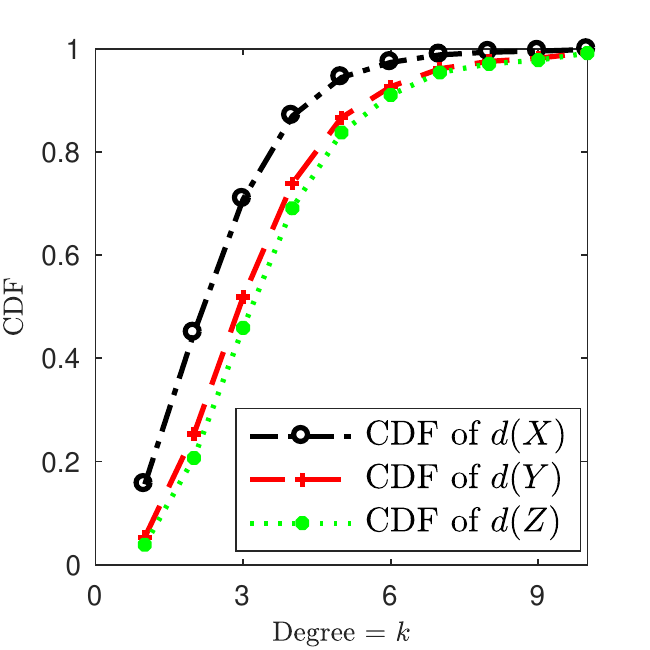}
		\caption{$r_{kk} = -0.2$ (disassortative network)}
		\label{subfig:fosdYZ_alpha2pt4_rkk_neg0pt2}
	\end{subfigure}\hfill
	\begin{subfigure}[!h]{0.3\textwidth}
		\centering
		\includegraphics[width=1.5in]{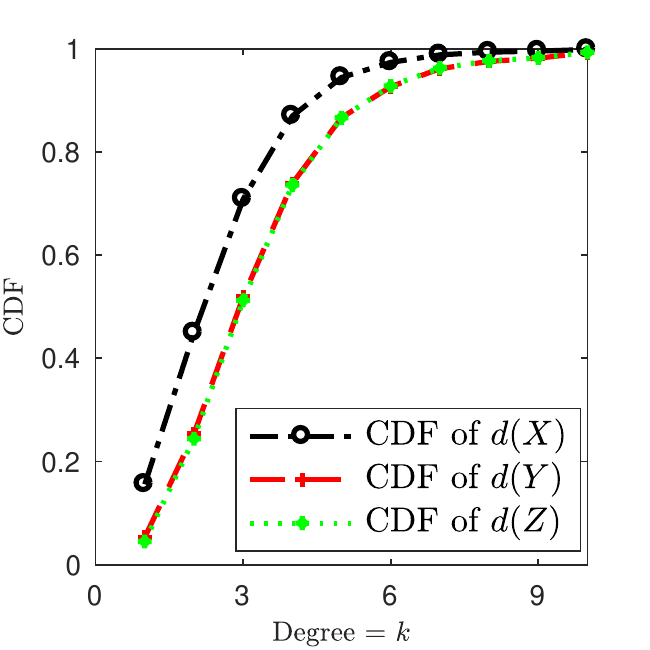}
		\caption{$r_{kk} = 0.0$}
		\label{subfig:fosdYZ_alpha2pt4_rkk_0}
	\end{subfigure}\hfill 
	\begin{subfigure}[!h]{0.3\textwidth}
		\centering
		\includegraphics[width=1.5in] {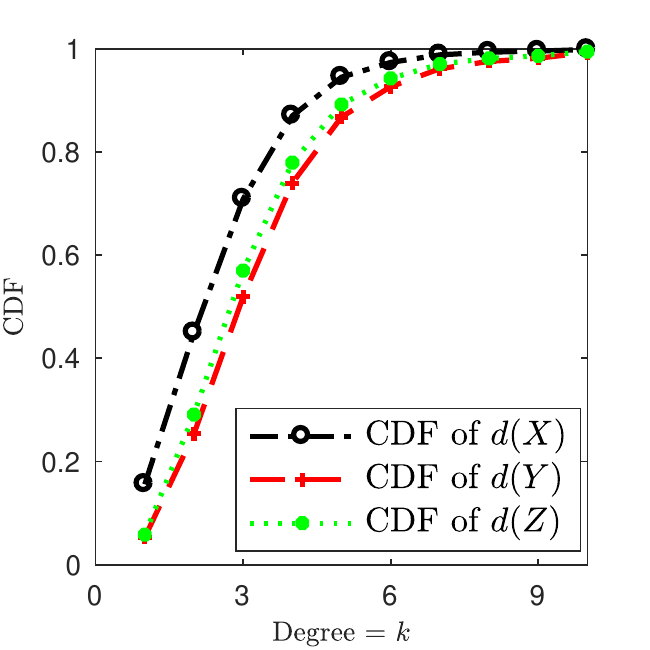}
		\caption{$r_{kk} = 0.2$ (assortative network)}
		\label{subfig:fosdYZ_alpha2pt4_rkk_0pt2}
	\end{subfigure}
	\caption{The cumulative distribution functions (CDF) of the degrees $d(X), d(Y), d(Z)$ of a random node ($X$), a random friend ($Y$) and a random friend ($Z$) of a random node respectively, for three graphs with the same degree distribution (power-law distribution with a coefficient $\alpha = 2.4$) but different neighbor-degree correlation coefficients $r_{kk}$, generated using the Newman's edge rewiring procedure. This illustrates that $\mathbb{E} \{d(Z)\} \geq \mathbb{E}\{d(Y)\}$ for $r_{kk} \leq 0$ (Fig. \ref{subfig:fosdYZ_alpha2pt4_rkk_neg0pt2}) and vice-versa. Further, this figure also shows how the distributions of $d(X), d(Y)$ remain invariant to the changes in the joint degree distribution $e(k,k')$ that preserve the degree distribution $P(k)$.}
	\label{fig:fosdYZ_alpha2pt4} 
\end{figure*}

\section{Non-Linear Bayesian Filtering for Estimating Population State}
\label{sec: NL_Filtering}
In Sec. \ref{sec:FP_based_polling}, algorithms to estimate the fraction of infected (state $1$) individuals in the case of slow diffusion dynamics where node states can be treated as fixed for the estimation purpose were discussed. However, treating states of nodes as fixed is not realistic when the diffusion takes place on the same time scale as the one on which individuals are polled (i.e. measurements are collected). Further, the non-linear dynamics (\ref{eq:MFD_approximation_X}) of the information diffusion rules out the possibility of applying standard Bayesian filtering methods such as Kalman filter to recursively update the population state estimate with new measurements \cite{krishnamurthy2016}. This section presents the non-linear Bayesian filtering method proposed in \cite{krishnamurthy2017tracking} which computes an optimal (in a mean-squared error sense) estimate of the population state with each new measurement. The method consists of two steps at each time instant: 
\begin{enumerate}
	\item sampling nodes from the network to obtain a noisy estimate of the population state 
	
	\item using the noisy estimate to compute the posterior distribution and then compute the new conditional mean of the estimate.
\end{enumerate}

\subsection{Sampling}{\label{subsec:samp}}

We first consider sampling the social network $G = (V,E)$ discribed in Sec. \ref{subsec:SIS_model} for the purpose collecting measurements to estimate the population state $\bst_{\dtime}$ at time $n$. We assume that the degree distribution $\degdist(\cdot)$ of the underlying network is known. Note that friendship paradox based NEP algorithms (presented in Sec. \ref{sec:FP_based_polling} for estimating the scalar valued fraction of infected nodes) can be easily extended for obtaining such (noisy) measurements of population state vector $\bst_{\dtime}$. For example, at each time instant $\dtime$, a random friend can be sampled and asked to provide an estimate of the population state $\bst_{\dtime}$ based on her neighbors. Apart from such extensions of  friendship paradox based NEP methods, we discuss two other widely used methods for sampling large networks for the purpose of obtaining an empirical estimate of $\bst_{\dtime}$.
\subsubsection{Uniform Sampling}
At each time $\dtime$,   $\sample(\degg)$ individuals are
sampled\footnote{For large population where $\vertexnum(d)$ is large, sampling with and without replacement are equivalent.}  independently and uniformly  from
the population $\vertexnum(\degg)$ comprising of agents with degree $\degg$.
Thus, a uniformly distributed independent sequence of nodes, denoted by $\{\nodem_{\seq}, \seq \in \{1,2,\hdots,\sample(\degg) \} \}$, is generated from the population  $\vertexnum(\degg)$.
From these independent samples, the empirical infected population state $\nodeobs_{\dtime}(\degg)$ of degree $\degg$ nodes at each time $\dtime $ is 
\beq \nodeobs_{\dtime}(\degg) = \frac{1}{\sample(\degg) }  \sum_{
	\seq=1}^{\sample(\degg)} \mathds{1} ( s_{\dtime}^{(\nodem_{\seq})}=1). \label{eq:sentiment} \eeq
At each time $\dtime$, $\nodeobs_{\dtime}$ can be viewed as noisy observation of the infected population state $\bst_{\dtime}$.

\subsubsection{MCMC Based Respondent-Driven Sampling (RDS)} \label{subsec:RDS}
Respondent-driven sampling~(RDS) was  introduced by Heckathorn~\cite{heckathorn1997respondent,heckathorn2002respondent} as an approach for sampling from hidden populations
in social networks and  has gained enormous popularity in recent years. In RDS sampling, current sample members recruit future sample members. 
The RDS procedure is as follows:  A small number of people in the target population serve as seeds. After participating in the study, the seeds recruit other people they know through the social network in the target population. The sampling continues according to this procedure  with current sample members recruiting the next wave of sample members until the desired sampling size is reached. 

RDS can be viewed as a form of Markov Chain Monte Carlo~(MCMC) sampling.
Let $\{\nodem_{\seq}, \seq \in \{1,2,\hdots,\sample(\degg) \} \}$  be the realization of an aperiodic irreducible Markov chain with
state space  $\vertexnum(\degg)$ comprising of nodes
of degree $\degg$. This Markov chain models the individuals of degree $\degg$ that are sampled, namely, the first individual $\nodem_{1}$ is sampled and then recruits the second
individual $\nodem_{2}$ to be sampled, who then recruits $\nodem_{3}$ and so on.
Instead  of the independent sample estimator~(\ref{eq:sentiment}),
an asymptotically unbiased MCMC estimate is computed as
\beq \frac{ \sum_{\seq = 1}^{\sample(\degg)}  \frac{\mathds{1} (s_{\dtime}^{(\nodem_{\seq})}=1)}{\steady{(\nodem_{\seq}})} }{ \sum_{\seq=1}^{\sample{(\degg)}}  \frac{1}
	{\steady{(\nodem_{\seq}})}}
\label{eq:mcmcrds}
\eeq
where  $\steady(\nodem)$, $\nodem \in \vertexnum(\degg)$, denotes the stationary distribution of the Markov chain $\nodem_\seq$.

In RDS, the transition matrix  and, hence, the stationary distribution $\steady$
in  the estimate~(\ref{eq:mcmcrds})
is specified as follows: Assume that  edges between any two nodes $i$ and $j$ have symmetric weights $\weight_{ij}$
(i.e.,
$\weight_{ij} = \weight_{ji}$). Node  $i$ recruits node $j$ with transition probability
$\weight_{ij}/ \sum_{j} \weight_{ij}$. Then, it can be easily seen that
the stationary distribution is
$\pi(i) = \sum_{j \in \Vertexset} \weight_{ij}/ \sum_{i \in \Vertexset, j \in \Vertexset} \weight_{ij}$. Using this stationary
distribution along with (\ref{eq:mcmcrds}) yields the RDS algorithm.
Since a Markov chain over a non-bipartite connected undirected network is aperiodic, the initial seed for  RDS can be picked arbitrarily, and the estimate~(\ref{eq:mcmcrds}) is asymptotically unbiased \cite{goel2009respondent}. 

The key outcome of this subsection is that by the central limit theorem (for an irreducible aperiodic finite state Markov chain), the
estimate of the probability that a node is infected in a large population (given its degree) is asymptotically Gaussian. For a sufficiently large number of samples, observation of the infected population state is approximately Gaussian, and  the sample observations can be expressed as 
\begin{equation} \label{eq:obs_inf}
	\obs_{n} = C \bst_n + v_n
\end{equation}
where $v_n \sim \mathscr{N}(\mathbf{0}, \mathbf{R})$ is the observation noise with the covariance matrix $\mathbf{R}$ and observation matrix $C$ dependent on the sampling process and $\bst_n \in \mathbb{R}^{\Ndeg}$ is the infected population state and evolves according to the polynomial dynamics (\ref{eq:MFD_approximation_X}).

\subsection{Non-linear filter and PCRLB for Bayesian Tracking of Infected Populations}{\label{sec:nl_fil}}
In Sec. \ref{sec:preliminaries}, the mean field dynamics for the population state as a system with polynomial dynamics (\ref{eq:MFD_approximation_X}) was discussed. Linear Gaussian observations (\ref{eq:obs_inf}) can be obtained by sampling the network as outlined in Sec. \ref{subsec:samp}. In this subsection, we consider Bayesian filtering for recursively estimating the infected population state $\bst_n$ in large networks. Then  posterior Cram\'er-Rao lower bounds (PCRLB) are obtained for these estimates.
We first describe how to express the mean field dynamics (\ref{eq:MFD_approximation_X}) in a form amenable to employing the non-linear filter described in \cite{HB14}.


\subsubsection{Mean Field Polynomial Dynamics}
Consider a $\Ndeg$-dimensional polynomial vector $f(\bst) \in \mathbb{R}^{\Ndeg}$:
\begin{equation}
f(\bst) = A_0 + A_1 \bst + A_2 \bst \bst^\prime + A_3 \bst \bst \bst^\prime + \dots
\label{eq:polynomialfunction}
\end{equation}
where the co-coefficients $A_0,A_1,\hdots,A_i$ are dimension $1,2,\hdots,(i+1)$ tensors, respectively.  Note that $A_i \bst \bst \dots \bst^\prime$ is a vector with $r^{\text{th}}$ entry given by
\begin{equation*}
A_i \bst \bst \dots \bst^\prime (r) = \sum_{j_1, j_2, j_3, \dots, j_i} A_i(r, j_1, j_2, \dots, j_i) \bst_{j_1} \bst_{j_2} \dots \bst_{j_i}
\end{equation*} 
where $A_i(r, j_1, j_2, \dots, j_i)$ is the $r, j_1, j_2, \dots, j_i$ entry of tensor $A_{i}$ and $x_{j}$ is the $j^{th}$ entry of $x$.
Because (\ref{eq:MFD_approximation_X}) has polynomial dynamics, it can be expressed in the form of (\ref{eq:polynomialfunction}) by constructing the tensors $A_i$. We refer the reader to \cite{krishnamurthy2017tracking} for the exact forms of these equations.

\subsubsection{Optimal Filter for Polynomial Dynamics}
With the mean-field dynamics (\ref{eq:MFD_approximation_X}) expressed in the form (\ref{eq:polynomialfunction}), we are now ready to describe the optimal filter to estimate the infected population state.
Optimal Bayesian filtering refers to recursively computing the conditional density (posterior) $p(x_n |Y_n)$, for $n=1,2,\cdots,$ where $Y_n$ denotes the observation sequence $y_1,\ldots, y_n$. From this posterior density, the conditional mean estimate $\mathbb{E}\{x_n |Y_n\}$ can be computed by integration. (The term optimal refers to the fact that the conditional mean estimate is the minimum variance estimate).  In general for nonlinear or non-Gaussian systems, there is no finite dimensional filtering algorithm, that is, the posterior  $p(x_n |Y_n)$ does not have a finite dimensional statistic. However, it is shown in \cite{HB14} that for Gaussian systems with polynomial dynamics, one can devise a finite dimensional filter (based on the Kalman filter) to compute the conditional mean estimate. That is, Bayes rule can be implemented exactly (without numerical approximation) to compute the posterior, and the conditional mean can be computed from the posterior. Therefore, to estimate the infected population state using the sampled observations (\ref{eq:obs_inf}), we employ this optimal filter.

The non-linear filter prediction and update equations are given as:\\
\underline{\textit{Prediction step}:} \\
\begin{align}\label{eq:filt_predict}
\hat{\bst}^-_{n} &= \mathbb{E}\{\bst_{n}|Y_{n-1}\} = \mathbb{E}\{f(\bst_{n-1})|Y_{n-1}\}\\ \nonumber
\varmi &= \mathbb{E}\{(\bst_n -\hat{\bst}_{n})(\bst_n -\hat{\bst}_{n})^\prime | Y_{n-1}\} \\ \nonumber
&=\mathbb{E}\{(f(x_{n-1}) - \mathbb{E}\{ f(x_{n-1}) |Y_{n-1}\} + v_{n-1}) \\ &\quad \quad \times(f(x_{n-1}) - \mathbb{E}\{ f(x_{n-1}) \nonumber |Y_{n-1}\} + v_{n-1})^\prime|Y_{n-1} \} \\ \nonumber
&= \mathbb{E}\{f(x_{n-1}) f(x_{n-1})' |Y_{n-1}\} - \mathbb{E}\{f(x_{n-1}) |Y_{n-1}\}\\ &\quad \quad \times \mathbb{E}\{f(x_{n-1}) |Y_{n-1}\}^\prime + \mathbf{Q}_{n-1} \nonumber
\end{align}
where $Y_n = \{ Y_{n-1},y_n\}$ denotes the observation process; $\varmi$ denotes the priori state co-variance estimate at time $\dtime$; and $v_n$ denotes the Gaussian state noise at time $n$, with covariance~$\mathbf{Q}_n$.  

The filter is initialized with mean $\hat{\bst}_0$ and covariance $H_0^{-}$.  The filter relies upon being able to compute the expectation  $\mathbb{E}\{f(\bst_{n-1}) f^\prime(\bst_{n-1})|Y_{n-1}\}$ in terms of $\hat{\bst}_{n-1}$ and $\varmi$. When $f(\cdot)$ is a polynomial, $f(\bst_{n-1}) f(\bst_{n-1})' $ is a function of $\bst_{n-1}$, and the conditional expectations in (\ref{eq:filt_update}) can be expressed only in terms of $\hat{\bst}_{n-1}$ and $\varmi$, permitting a  closed form{\footnote{For an explicit implementation of such a filter for a third order system with an exact priori update equation for $\varmi$ and $\hat{\bst}_{n}^-$, see \cite{HB14}.}} prediction step. \\
\underline{\textit{Update step}:}  \\
\begin{align} \label{eq:filt_update}
\hat{\bst}_{n} &= \mathbb{E}\{\bst_{n}|Y_{n}\}= \hat{\bst}^-_{n} +\varmi C^\prime(\rn +C\varmi C^\prime)^{-1}(\obs_{n} -C \hat{\bst}^- _{n}) \nonumber \\
\gn &= \varmi C^\prime(\rn +C \varmi C^\prime)^{-1} \nonumber \\
\varp &= (I-\gn C) \varmi (I - \gn C)^\prime + \gn \rn \gn^{\prime} 
\end{align} 
where $\hat{\bst}_n$ denotes the conditional mean estimate of the state and $\varp$ the associated conditional covariance at time $n$. $C$ denotes the state observation matrix; $\mathbf{R}_n$ denotes the observation noise co-variance matrix; $\gn$ denotes the filter gain; and $I$ denotes the identity matrix.

Since the dynamics of (\ref{eq:MFD_approximation_X}) are polynomial, the prediction and update steps of (\ref{eq:filt_predict}) and (\ref{eq:filt_update}) can be implemented without approximation. These expressions constitute the optimal non-linear filter and can be used to track the evolving infected population state.

\section{Summary and Discussion}
This chapter discussed in detail, three interrelated topics in information diffusion in social networks under the central theme of dynamic modeling and statistical inference of SIS models. 

Firstly, Sec. \ref{sec:effects_FP_SISModel} showed that the effect of high degree nodes updating their states (infected or susceptible) more frequently is reflected in the update term of the deterministic mean-field dynamics model and, does not affect the critical thresholds which decide if the information diffusion process will eventually die out or spread to a non-zero fraction of individuals. Secondly, the case where two-hop neighbors are influencing the evolution of states in the SIS model was discussed. This two-hop neighbor influence, called monophilic contagion, was shown to make the SIS information diffusion easier (by lowering the critical thresholds). Sec. \ref{sec:active_networks} extended the mean-field model to the case where the underlying social network randomly evolves depending on the state of the information diffusion. How the collective dynamics of such a process can be modeled by a deterministic ordinary differential equation with an algebraic constraint was discussed.

Related to the statistical inference aspect of the SIS information diffusion processes, how the state of the underlying population (induced by the SIS model) can be estimated was explored under two cases. Firstly, for the case where the dynamics of the SIS model is slower (and hence node states can be treated as fixed for estimation purpose) compared to measurement collection (polling), friendship paradox based  polling algorithms to estimate the fraction of infected nodes were discussed. Such algorithms can outperform classical polling methods such as intent polling by lowering the variance of the estimate. Secondly, for the case where the dynamics of the SIS model evolve on the same time scale as the measurement collection process, a non-linear Bayesian filtering algorithm which harvests the polynomial dynamics of the SIS model was discussed. This filtering algorithm is an optimal filter which updates the state estimate with each new measurement. 

\vspace{0.25cm}
\noindent
{\bf Future research directions: } The topics discussed in this chapter yield interesting directions for future research in information diffusion processes. Firstly, the nodes were treated as non-strategic decision makers in this chapter i.e. their decisions to update the states are not strategic. The changes in the dynamics and critical thresholds yielded by the case where nodes are strategic utility maximizers is an interesting direction for future research. Secondly, the topics of this chapter focused on the case where the underlying network is an undirected graph. The mean-field study of such topics (diffusion based on two-hop contagion, effects of friendship paradox and filtering) in the context of a directed graph remains an interesting research direction to be explored. There is also substantial motivation to evaluate SIS models using real data; some preliminary results applied to YouTube appear in \cite{HAK16,KH18}.

\vspace{0.25cm}
\noindent
{\bf Topics beyond the scope of the chapter: }Since the current chapter dealt with SIS contagions that are modeled using mean-field dynamics, several important topics related to the diffusion of information in social networks are beyond the scope of the current chapter. Some of such topics include: social learning \cite{krishnamurthy2012quickest,krishnamurthy2013social,chamley2004rational}, influence maximization \cite{kempe2003}, strategic agents and game theoretic learning \cite{gharehshiran2013,jackson2016friendship} and, inferring the network structure using diffusion traces \cite{gomez2012inferring}.


\bibliographystyle{spmpsci}
\bibliography{IMA_2018_ref_v4}

\end{document}